\documentclass[aps,physrev,twocolumn,superscriptaddress]{revtex4-2}
\usepackage{graphicx}
\usepackage{dcolumn}
\usepackage{bm}
\usepackage{multirow}
\usepackage{capt-of}
\usepackage{amsmath}
\usepackage{mathtools}
\usepackage[normalem]{ulem}
\usepackage{braket}

\begin{document}

\title{Simultaneous High-Fidelity Readout and Strong Coupling in a Donor-Based Spin Qubit}


\author{Si Yan Koh}
\author{Weifan Wu}
\author{Kelvin Onggadinata}
\author{Arghya Maity}
\affiliation{School of Physical and Mathematical Sciences, Nanyang Technological
University, 21 Nanyang Link, Singapore 637371, Singapore}

\author{Mark Chiyuan Ma}
\affiliation{Department of Physics, Faculty of Science, National University of Singapore, Singapore 117551, Singapore}

\author{Calvin Pei Yu Wong}
\affiliation{Institute of Materials Research and Engineering (IMRE), Agency for Science,
Technology and Research (A*STAR), Singapore 138634, Singapore}
\affiliation{Centre for Quantum Technologies, National University of Singapore, Singapore 117543, Singapore}

\author{Kuan Eng Johnson Goh}
\affiliation{School of Physical and Mathematical Sciences, Nanyang Technological
University, 21 Nanyang Link, Singapore 637371, Singapore}
\affiliation{Department of Physics, Faculty of Science, National University of Singapore, Singapore 117551, Singapore}
\affiliation{Institute of Materials Research and Engineering (IMRE), Agency for Science,
Technology and Research (A*STAR), Singapore 138634, Singapore}
\affiliation{Centre for Quantum Technologies, National University of Singapore, Singapore 117543, Singapore}

\author{Bent Weber}
\affiliation{School of Physical and Mathematical Sciences, Nanyang Technological
University, 21 Nanyang Link, Singapore 637371, Singapore}

\author{Hui Khoon Ng}
\affiliation{Department of Physics, Faculty of Science, National University of Singapore, Singapore 117551, Singapore}
\affiliation{Centre for Quantum Technologies, National University of Singapore, Singapore 117543, Singapore}

\author{Teck Seng Koh}
\affiliation{School of Physical and Mathematical Sciences, Nanyang Technological
University, 21 Nanyang Link, Singapore 637371, Singapore}
\
\date{\today}

\begin{abstract}
Superconducting resonators coupled to solid-state qubits offer a scalable architecture for long-range entangling operations and fast, high-fidelity readout. Realizing this requires low photon-loss rates and qubits with tunable electric dipole moments that couple strongly to the resonator's electric field while maintaining long coherence times. For spin qubits, spin-photon coupling is typically achieved via spin-charge hybridization. However, this introduces a fundamental trade-off: a large spin-charge admixture enhances the coupling strength, which boosts readout and resonator-mediated gate speeds, but exposes the qubit to increased decoherence, thereby increasing the threshold required for strong coupling and limiting the time available for accurate state measurement. This makes it essential to identify optimal operating points for each qubit platform. We address this for the donor-based flip-flop qubit, whose microwave-controllable electron-nuclear spin states make it suitable for coupling to microwave resonators. We demonstrate that, by choosing intermediate tunnel couplings that balance strong interaction with long qubit lifetimes, high-fidelity readout and strong coupling are simultaneously achievable. We also map out the respective charge-photon couplings and photon-loss rates required. Furthermore, we show that experimental constraints on charge-photon coupling and photon loss can be mitigated using squeezed input fields. As similar trade-offs appear in quantum-dot-based qubits, our methods and insights extend naturally to these platforms, offering a potential route toward scalable architectures.
\end{abstract}

\maketitle

\section{\label{sec:intro}Introduction\protect}
Semiconductor spin qubits in silicon are strong candidates for quantum information processing \cite{kane1998,loss1998} due to their exceptionally long lifetimes and coherence times \cite{becher2023, chatterjee2021,watson2017,tyryshkin2012,saeedi2013, morello2020,stano2022, steinacker2025,yang2013}. These advantages are particularly pronounced in isotopically purified silicon, which gives a spin-free environment. Furthermore, silicon's compatibility with the conventional electronics industry facilitates integration into existing manufacturing processes, while the qubit's nanoscale footprint offers a promising pathway for scalability.

A potential challenge, however, is the limited range of the exchange interaction between spins \cite{asenov2003,morello2020}, which restricts inter-qubit connectivity. To overcome this, proposals have adapted the circuit quantum electrodynamics (cQED) architecture, where microwave photons in resonant cavities mediate interactions between distant qubits. This can be achieved with a modular architecture comprising well-characterized qubit arrays coupled via transmission line resonators \cite{vandersypen2017}. These resonators not only mediate interactions but also facilitate quantum non-demolition readout \cite{blais2004,childress2004,burkard2006,trif2008,vandersypen2017,xiang2013,srinivasa2024}. A key requirement for this scheme is a tunable electric dipole moment of the qubit capable of coupling strongly to the resonator's electric field, as direct magnetic coupling is typically too weak to be practical \cite{chatterjee2021}.

Donor- and quantum-dot-based platforms typically satisfy this requirement by employing an engineered or intrinsic spin-charge hybridization mechanism \cite{osika2022,krauth2022,tosi2017,petersson2012}. This mixing endows the spin with an effective electric dipole moment, inducing an effective spin-photon coupling. Strong coupling is achieved when this interaction exceeds both the resonator loss and qubit decoherence rates. However, hybridization introduces a fundamental trade-off: increasing the spin-charge admixture enhances the dipole moment, which boosts spin-photon coupling, crucially giving faster resonator-mediated gates, and measurement rates for qubit readout. A faster measurement improves readout fidelity by increasing the signal-to-noise ratio (SNR) and allowing the state to be resolved before the qubit relaxes. On the other hand, the same admixture increases the qubit's exposure to electrical noise and relaxation channels, thereby increasing the decoherence threshold required for strong coupling and shortening the qubit lifetime. Therefore, parameters that maximize coupling strength may inadvertently compromise readout fidelity. This competing effect necessitates identifying optimal operating points for each qubit platform to be viable, which naturally raises the questions: can strong coupling and high-fidelity readout be achieved simultaneously? If so, what system parameters are required? In this manuscript, we address these using the donor-based flip-flop qubit-resonator platform as a case study. 

In the donor-based flip-flop qubit, quantum information is encoded in electron-nuclear spin states, while the electric dipole moment is tuned by delocalizing the electron wavefunction between the donor and an interface (or a nearby donor) via an electric field \cite{tosi2017,truong2021,osika2022}. Theoretical estimates place the spin-photon coupling at approximately $2-3$~MHz \cite{tosi2017,osika2022}. Given typical resonator loss rates on the order of $1-10$ MHz, this system sits at the boundary between weak and strong coupling. Additionally, the flip-flop qubit's lifetime is reduced by up to eight orders of magnitude compared to donors in bulk silicon due to enhanced spin-valley relaxation \cite{boross2016}, potentially degrading its readout performance. These unique constraints make it interesting and important to determine how effectively the flip-flop qubit can leverage its advantages within the cQED architecture.

Our results are summarized as follows. We map the parameter space governing qubit-photon interactions and identify regimes where the conventional dispersive shift approximation holds, providing corrections where it breaks down. By rigorously accounting for critical photon numbers and parameter restrictions, we characterize the regions where strong coupling can be realized and compare them against regions of high readout performance. We quantify readout performance using the square of the SNR, where $\text{SNR}^2\ge 282$ corresponds to a single-shot readout fidelity of $F\ge99\%$. We find that performance is primarily dominated by the fraction of input photons that contribute to readout \cite{danjou2019}, which we call the efficiency.

Importantly, we find that regimes supporting simultaneous strong coupling and high-fidelity readout exist in principle, but their experimental realization is constrained by the achievable charge-photon coupling $g_c$ and photon-loss rate $\kappa$. These regimes are widest at intermediate tunnel couplings, which offer a broad window of favorable spin-photon coupling strengths and SNR. Finally, we show that readout performance can be further enhanced using squeezed input fields, which relax these constraints; we delineate the feasible limits of this technique. Because similar trade-offs and physical considerations appear in quantum-dot-based qubits, our methods and insights extend naturally to those platforms, offering a potential route toward scalable architectures.

The paper is structured as follows. In Sec.~\ref{sec:ff}, we introduce the theoretical qubit-resonator model and the dominant relaxation and decoherence processes of the flip-flop qubit. Sec.~\ref{sec:dis} details the theory and protocol for dispersive readout with the resonator. Here, we employ the well-established input-output theory and the quantum Langevin equation to derive the readout SNR. We detail the assumptions, restrictions, and approximations made in our calculations. Our main results and parameter maps are presented in Sec.~\ref{sec:readoutresults}. We conclude in Sec.~\ref{sec:conc}.
\section{\label{sec:ff}Theoretical Model}
The total Hamiltonian for the donor-based flip-flop qubit coupled to a resonator is:
\begin{eqnarray}
    \label{eq:fullH}
    H_{\text{tot}} &&= H_{\text{FF}} + H_{\text{r}} + H_{\text{int}}.
\end{eqnarray}
$H_{\text{FF}}$ is the flip-flop qubit Hamiltonian, $H_{\text{r}}$ is the resonator Hamiltonian, and $H_{\text{int}}$ is the qubit-resonator interaction Hamiltonian. In this work, we take $\hbar=1$. 
\subsection{Flip-flop qubit Hamiltonian}
\begin{figure}
{\includegraphics[width=0.50\textwidth]{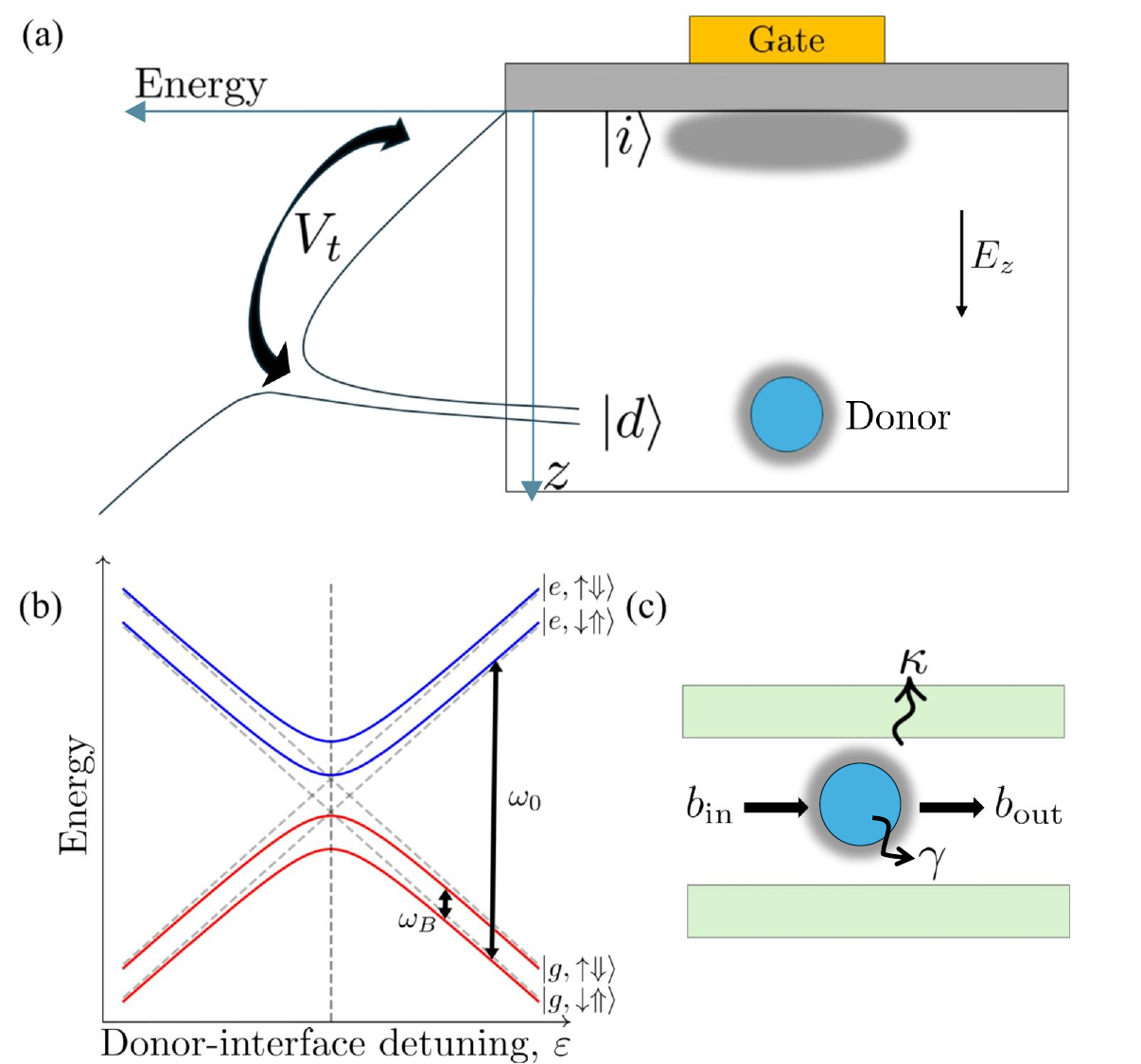}}
\caption{\label{fig:fig1} (a) Schematic of the flip-flop qubit. The electron wavefunction (grey) can be delocalized from the donor state towards the silicon-insulator interface with an electric field $E_z$ from a top gate. The donor and interface states are coupled by $V_t$. (b) Energy level diagram of the flip-flop qubit. The electron-nuclear spin states $|\uparrow\Downarrow\rangle,|\downarrow\Uparrow\rangle$ are separated approximately by the Zeeman splitting $\omega_B$, while the orbital splitting $\omega_0$ separates the ground (red) and excited (blue) charge manifold. The flip-flop qubit is encoded in the spin states of the ground manifold (red). (c) Schematic of dispersive readout.  The flip-flop qubit is coupled to the resonator of frequency $\omega_r$ with charge-photon coupling strength $g_c$. The qubit experiences an effective relaxation rate $\gamma$ and the resonator photon-loss rate is $\kappa$. $b_{\text{in}}$ and $b_{\text{out}}$ are the known input and measured output drives that are used to read the qubit state.}
\end{figure}

The qubit consists of a phosphorus donor comprising a spin-1/2 electron (\(\{\ket{\downarrow},\ket{\uparrow}\}\), \(\gamma_{e}=2\pi \times 29.97\ \mathrm{GHz\ T}^{-1}\)) and a spin-1/2 nucleus (\(\{\ket{\Downarrow},\ket{\Uparrow}\}\), \(\gamma_{n}=2\pi \times 17.23 \text{ MHz T}^{-1}\)) \cite{tosi2017} (Fig.~\ref{fig:fig1}(a)).

A top-gate electric field $E_z$ delocalizes the electron wavefunction from the donor to the silicon-insulator interface. This delocalization is parameterized by $\varepsilon$, where $\varepsilon=0$ corresponds to equal admixtures of donor- and interface-centered states, \(\{\ket{d},\ket{i}\}\). These states are tunnel coupled by $V_t$ and span the qubit charge basis, expressible in terms of excited and ground orbital eigenstates \(\ket{e}\) and \(\ket{g}\) with splitting $\omega_0$. Additionally, the hyperfine interaction \(A = 117 \text{ MHz}\) couples the electron and nuclear spins when the electron wavefunction overlaps with the donor. Together, the hyperfine interaction and electric field enable tunable spin-charge hybridization.

The qubit logical states are approximately the ground manifold electron-nuclear spin states, \(\ket{g,\uparrow\Downarrow}\) and \(\ket{g,\downarrow\Uparrow}\), whose separation is primarily set by the Zeeman splitting $\omega_B$ (Fig.~\ref{fig:fig1}(b)). This is provided the excited orbitals are well-separated ($\omega_0 > \omega_B$) and the states are predominantly spin-like ($|\omega_0-\omega_B|\gg A/4$)~\cite{truong2021,boross2016}.

The flip-flop Hamiltonian in the tensor product of charge and Zeeman eigenbases is \cite{tosi2017,truong2021}:
\begin{eqnarray}
    \label{eqn:ffH}
    H_{\text{FF}}&&=H_{\text{charge}} + H_{\text{B}} + \Delta H_{\text{B}} + H_{\text{A}} 
\end{eqnarray}
where 
\begin{eqnarray}
    H_{\text{charge}} &&= -\frac{1}{2}\omega_0\tau_z  \\
    H_{\text{B}} &&= -\frac{1}{2}\omega_B\sigma_z \\
    \Delta H_{\text{B}} &&=-\frac{1}{4}\Delta \omega_{B}(1+\cos\eta \tau_{z} + \sin\eta \tau_{x})\sigma_{z}\\
    H_{\text{A}} &&=-\frac{1}{8}A(1-\cos\eta \tau_{z} - \sin\eta \tau_{x})(1-2\sigma_{x})
\end{eqnarray}
Here, $H_{\text{charge}}$ describes the charge eigenstates $\ket{e} = \sin\left(\eta/2\right)\ket{i}+\cos\left(\eta/2\right)\ket{d}$ and $\ket{g} = \cos\left(\eta/2\right)\ket{i}-\sin\left(\eta/2\right)\ket{d}$, with orbital admixture angle ${\eta} = \arctan(V_t/\varepsilon)$ and $\tau_z \equiv \ket{g}\bra{g} - \ket{e}\bra{e}$. $H_{\text{B}}$ is the Zeeman term with $\sigma_z \equiv \ket{\downarrow \Uparrow}\bra{\downarrow \Uparrow} - \ket{\uparrow \Downarrow}\bra{\uparrow \Downarrow}$. Spin-charge hybridization arises primarily from the site-dependent hyperfine interaction $H_{\text{A}}$ and a gyromagnetic ratio correction $\Delta H_{\text{B}}$ at the Si/SiO$_2$ interface. Atomistic calculations \cite{rahman2009,truong2021} indicate this correction is small: $|\Delta\omega_{B}/\omega_B| \lesssim 0.7\%$. We choose $\Delta\omega_B = -0.2\%\omega_B$ in this work~\cite{tosi2017,boross2016}.
\subsection{Qubit-resonator interaction}

The qubit electric dipole couples to the resonator with strength $g_c$. Combined with spin-orbit mixing, this enables dispersive readout of the spin state (Fig.~\ref{fig:fig1}(c)) and effective spin-photon coupling. The resonator Hamiltonian ($H_{\text{r}}$) and interaction Hamiltonian ($H_{\text{int}}$) \cite{warren2019,beaudoin2016}, in the charge eigenbasis, are:
\begin{eqnarray}
    H_{\text{r}} &&= \omega_ra^{\dagger}a \\
    H_{\text{int}} &&= g_c\left(\sin\eta\tau_x + \cos\eta\tau_z\right)\left(a+a^{\dagger}\right).
\end{eqnarray}

\subsection{\label{sec:disH}Effective Hamiltonian in dispersive regime}
In the dispersive regime with large qubit-resonator detuning, the interaction induces a qubit-state-dependent resonator frequency shift, enabling dispersive readout. Assuming well-separated excited orbitals ($\omega_0 > \omega_B, \omega_r$) and strongly spin-like qubit states ($|\omega_0-\omega_B|\gg A/4$), the relevant dynamics are confined to the ground orbitals. We apply the Schrieffer-Wolff transformation to account for virtual effects of higher orbitals \cite{schrieffer1966,bravyi2011}, projecting the effective Hamiltonian onto the lower-energy subspace. If energy assumptions fail, e.g., the perturbation $g_c$ is comparable to the smallest transition energies, neglected couplings may open spin relaxation channels, degrading readout \cite{danjou2019,srinivasa2013}.

Decomposing the total Hamiltonian into terms that do and do not couple between the charge, spin, or resonator subspaces allows us to treat coupling terms as perturbations ($V = V_{\text{od}}+V_{\text{d}}$)~\cite{warren2019}, with validity conditions discussed in Sec.~\ref{sec:para_res}. $H_{\text{tot}}$ consists of:
\begin{eqnarray}
H_0 = &&\left(\frac{1}{8}A\cos{\eta}-\frac{1}{2}\omega_0\right)\tau_z+\frac{1}{8}A\sin{\eta}\tau_{x}\nonumber\\
&&+\left(-\frac{1}{2}\omega_B-\frac{1}{4}\Delta\omega_B\right)\sigma_z+\frac{1}{4}A\sigma_x\nonumber \\
&&-\frac{1}{8}A+\omega_ra^{\dagger}a\\
\label{eq:Vod2}
 V_{\text{od}} =&&-\frac{1}{4}\Delta\omega_{B}\sin{\eta}\tau_{x}\sigma_{z}-\frac{1}{4}A\sin{\eta}\tau_{x}\sigma_{x}\nonumber \\
 &&+g_{c}\sin{\eta}\tau_{x}(a+a^{\dagger})  \\
  V_{\text{d}}=&&-\frac{1}{4}\Delta\omega_{B}\cos{\eta}\tau_{z}\sigma_{z}-\frac{1}{4}A\cos{\eta}\tau_{z}\sigma_{x}\nonumber\\
  &&+g_{c}\cos{\eta}\tau_{z}(a+a^{\dagger})
\end{eqnarray}
We perform two Schrieffer-Wolff transformations (full details in Appendix~\ref{sec:SWderivation}). The first perturbatively eliminates block off-diagonal terms responsible for orbital transitions, rendering the Hamiltonian block-diagonal to first order in $V$. This yields the effective spin-photon coupling and slightly rotates the charge basis, giving us new $V_{\text{od}}$ and $V_{\text{d}}$ for the second transformation. To preserve all interactions to second order, we do not project onto the orbital ground state at this stage; such a projection would only affect results at higher orders. The second transformation decouples spin-flip transitions, yielding the effective Hamiltonian in the dispersive regime:
\begin{eqnarray}
    H_{\text{eff}} &&= H_{\text{dis}} + H_{\text{tr}}
\end{eqnarray}
where
\begin{eqnarray}
    H_{\text{dis}} =&& -\frac{1}{2}\tilde{\omega}_{B} \sigma_z + \left(\tilde{\omega}_{r}  + \chi_{z}\sigma_z\right)a^{\dagger}a.
\end{eqnarray}
The tilde indicates effective terms post-transformation (see Appendix~\ref{sec:SWderivation}). Provided the coupling strength in $H_{\text{tr}}$ is small compared to the minimum energy gap, these terms are negligible. Importantly, $H_{\text{dis}}$ includes the dispersive shift $\chi_z$, which underpins readout fidelity by controlling the distinguishability of the two qubit spin states, given by
\begin{eqnarray}
\label{eq:chi_1}
    \chi_z =&& g_{s}^{2}\left(-\frac{1}{\tilde{\Delta}_-}-\frac{1}{\tilde{\Delta}_+}\right)+\chi_{\text{cor}}
\end{eqnarray}
where $\tilde{\Delta}_\pm$ is defined in Appendix~\ref{sec:SWderivation} (Eq.~\eqref{eq:tildedefinitions}). $\chi_{\text{cor}}$ represents corrections due to additional couplings that are generally negligible in our parameter space (Sec.~\ref{sec:workregime}). $g_s$ denotes the spin-photon coupling strength associated with spin-flip transitions within each orbital (the $\tau_z\sigma_x\left(a+a^{\dagger}\right)$ term).

\subsection{\label{sec:qubitrelax}Decoherence and relaxation processes}
Readout is sensitive to qubit relaxation and photon losses. In bulk silicon, phosphorus donor spin relaxation via phonon emission through the electron-phonon and spin-orbit couplings is negligible ($\sim 10^{-4}$~Hz) at low temperatures \cite{pines1957}. However, gate-induced electric fields concentrate the interface wavefunction into specific valleys, instead of evenly distributing them across all six valleys, enhancing this rate by up to 8 orders of magnitude \cite{baena2012,kohn1955,zwanenburg2013}. Fermi's Golden Rule yields the zero-temperature relaxation rate to leading order in the small parameter $A/\left(\omega_0-\omega_B\right)$~\cite{boross2016}:
\begin{eqnarray}~\label{eq:gamma_FF}
    \gamma_{\text{FF}} &&= \frac{1}{4}\frac{A^{2}V_{t}^{2}\omega_{B}^{3}}{\omega_{0}^{3}\left(\omega_{0}^{2}-\omega_{B}^{2}\right)^{2}}\gamma_{0},
\end{eqnarray}
where $\gamma_{0}= \Theta\omega_0V_t^2$ is the orbital relaxation rate. $\Theta = 2.37 \times 10^{-24}~\text{s}^2$ depends on silicon material properties \cite{tosi2017,boross2016} like uniaxial deformation potential, mass density, and longitudinal and transverse sound velocity. Numerical verification confirms this expression remains valid in the dispersive regime. Since $\omega_0 > \omega_B$, we avoid the relaxation hot spot (typically $\sim 10$ kHz) that occurs when they are resonant \cite{boross2016}, and $\gamma_{\text{FF}}$ decreases with increased $\omega_0$. At finite temperatures, $\gamma_{\text{FF}}$ is modified by the Bose-Einstein factor. Other relaxation processes, such as electron spin relaxation, are negligible \cite{boross2016,tosi2017}.

We account for photon loss via the Purcell decay. When the magnitude of the spin-resonator detuning greatly exceeds the photon-loss rate ($|\Delta_-| \equiv |\omega_{B}-\omega_r|\gg\kappa$), which is expected in the dispersive regime, the rate is approximately
\begin{equation}\label{eq:purcell_rate}
    \gamma_{\text{pu}} \approx \frac{g_{s}^{2}\kappa}{\Delta_-^{2}}.
\end{equation}
(See Sec.~\ref{sec:workregime} for further justification). As these processes are independent, and depending on the specific values of $\kappa$ and $g_c$, one process does not always dominate the other, the total relaxation rate is simply a sum of the two:
\begin{eqnarray}
    \gamma = \gamma_{\text{FF}} + \gamma_{\text{pu}}
\end{eqnarray}

We assess the strong coupling regime by comparing spin-photon coupling strength to decoherence and photon-loss rates. For quantum dot and donor spin qubits, coherence is often limited by dephasing~\cite{hu2025,muhonen2014,thorvaldson2025,pla2013,tejel2021,veldhorst2014, ten2019}. In particular, we account for the effect of $1/f$ quasi-static charge noise, a known major source of decoherence in solid-state systems~\cite{truong2021,dutta1981,paladino2014,sanjose2006,connors2019}. We model this as Gaussian, where qubit frequency depends on noise parameter $\delta\varepsilon$:
\begin{eqnarray}
    \tilde{\omega}_B\left(\varepsilon + \delta\varepsilon\right) \approx \tilde{\omega}_B\left(\varepsilon\right) + \left|\frac{\partial\tilde{\omega}_B}{\partial\varepsilon}\right|\delta\varepsilon = \tilde{\omega}_B\left(\varepsilon\right) + \delta\omega
\end{eqnarray}
The qubit accumulates random phase $\phi(t) = \delta\omega t$. For a Gaussian decay function $C(t)$, the dephasing rate $\gamma_\phi$ is
\begin{eqnarray}
    C(t) &&= \langle e^{i\phi(t)}\rangle = e^{-\sigma_\phi^2 t^2/2} = e^{-{\gamma_\phi^2}t^2}\\
    \gamma_\phi &&= \frac{1}{\sqrt{2}}\left|\frac{\partial\tilde{\omega}_B}{\partial\varepsilon}\right|\delta\varepsilon
\end{eqnarray}
In this paper, we assume an RMS noise of $100\ \text{V}\ \text{m}^{-1}$, corresponding to $\delta\varepsilon \approx 0.363\ \text{GHz}$ for a $d= 15\ \text{nm}$ donor depth, consistent with Refs.~\cite{tosi2017,truong2021}. The overall decoherence rate is thus
\begin{eqnarray}
    \gamma_{\text{dec}} = \frac{\gamma}{2} + \gamma_{\phi}
\end{eqnarray}
When working at $\varepsilon = 0$, away from the flip-flop qubit sweet spots (Sec.~\ref{sec:sweetspot}), $\gamma_\phi \gg \gamma$, implying $\gamma_{\text{dec}}\approx\gamma_{\phi}$. This decoherence drops exponentially with higher $V_t$. 

\section{\label{sec:dis}Dispersive Readout}
\subsection{Input-output theory}
In this section, we show the figures of merit for readout performance using input-output theory and the quantum Langevin equation \cite{gardiner1985,benito2017}. These frameworks map the input drive $b_{\text{in}}(t)$ to the output signal $b_{\text{out}}(t)$, measured with homodyne detection, via the resonator field $a(t)$, allowing the qubit state to be inferred.

\begin{figure}
{\includegraphics[width=0.5\textwidth]{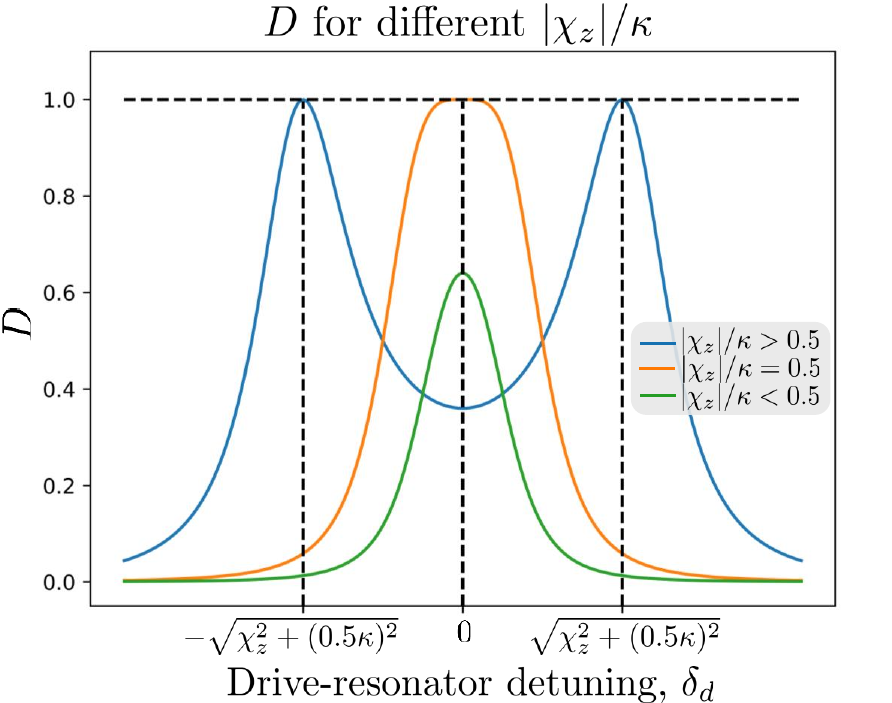}}
\caption{\label{fig:fig2} Readout efficiency function $D$ for different $|\chi_z|/\kappa$ regimes. For $|\chi_z|/\kappa \leq 0.5$, the optimal driving frequency is on resonance ($\delta_d = 0$, $\omega_d = \tilde{\omega_r}$), though $D<1$ for $|\chi_z|/\kappa < 0.5$. Conversely, when $|\chi_z|/\kappa > 0.5$, $D$ reaches unity at $\delta_d = \pm \sqrt{\chi_z^2 - \left(0.5\kappa\right)^2}$.}
\end{figure}

The resonator dynamics follow the quantum Langevin equation:
\begin{eqnarray}
    \label{eq:qle}
    \dot{a}(t) = i\left[H_{\text{dis}},a(t)\right] -\frac{\kappa}{2}a(t) + \sqrt{\kappa}b_{\text{in}}(t),
\end{eqnarray}
related to the input and output fields by \cite{gardiner1985}:
\begin{eqnarray}
    \label{eq:io}
    b_{\text{out}}(t) - b_{\text{in}}(t) = \sqrt{\kappa}a(t).
\end{eqnarray}
Typically, a coherent drive is applied to the resonator, which may carry some noise, so we can make the replacement $b_{\text{in}}(t) \rightarrow b_{\text{in}}(t) + \beta_{\text{in}}(t)$ \cite{danjou2019, blais2021}, where $b_{\text{in}}(t)$ is some input noise, typically taken to be Gaussian and white. $\beta_{\text{in}}$ is the input drive, and the resulting term $\sqrt{\kappa}\beta_{\text{in}}$ from Eq.~\eqref{eq:io} can be absorbed into the Hamiltonian with the replacement $H_{\text{dis}}\rightarrow H_{\text{dis}} + H_{\text{drive}}(t)$ \cite{blais2021} where
\begin{eqnarray}
    H_{\text{drive}}(t) = i\sqrt{\kappa}\Omega(t)\left(a^{\dagger}e^{-i\omega_dt-i\theta_d} - a e^{i\omega_dt+i\theta_d}\right).
\end{eqnarray}
and $\Omega(t)$ is the drive amplitude, while $\theta_d$ is the drive phase. The quantum Langevin equation can be updated to yield
\begin{eqnarray}
    \dot{a}(t) =& i\left[H_{\text{dis}} + H_{\text{drive}}(t), a(t)\right] -\frac{\kappa}{2}a(t) + \sqrt{\kappa} b_{\text{in}}(t) 
\end{eqnarray}
Moving to the interaction picture rotating with the drive $U_0(t) = e^{-i\omega_da^{\dagger}at}$ under the rotating wave approximation (RWA) yields:
\begin{eqnarray}
    \dot{a}(t) = &&-i\left(\delta_d + \chi_z\sigma_z\right)a(t)-\sqrt{\kappa}\Omega(t)e^{-i\theta_d}\nonumber \\
    &&-\frac{\kappa}{2}a(t)+\sqrt{\kappa}b_{in}(t)
\end{eqnarray}
where $\delta_d = \tilde{\omega_r} - \omega_d$ is the resonator-drive detuning. Following Ref.~\cite{danjou2019}, we assume a steady state and distinct input/output ports with equal photon-loss rates ($\kappa_{\text{in}} = \kappa_{\text{out}} = \kappa/2$) such that $\langle b_{\text{in}}(t)\rangle = 0  $ and $\beta_{\text{in}}(t)-\beta_{\text{out}}(t) = \sqrt{\kappa}a(t)$. The output signal expectation values for the qubit states ($\pm$) are thus:
\begin{eqnarray}
    \braket{\beta_{\text{out}}^{\pm}} =&& \frac{\sqrt{\kappa_{\text{in}}\kappa_{\text{out}}}\langle \Omega \rangle }{-i\left(\delta_d \pm \chi_z\right)-\frac{\kappa}{2}} = \frac{\sqrt{\kappa^2/4}\langle \Omega \rangle }{-i\left(\delta_d \pm \chi_z\right)-\frac{\kappa}{2}},
\end{eqnarray}
The steady-state readout contrast, which quantifies the distinguishability of the qubit states through the frequency shift, is
\begin{eqnarray}
    \left|\Delta\beta_{\text{out}}\right|^2 :=&& \left|\braket{\beta_{\text{out}}^{+}} - \braket{\beta_{\text{out}}^{-}}\right|^2 = \frac{\kappa}{2}\langle n\rangle D
\end{eqnarray}
Here, $\langle n \rangle$ is the mean photon number. The function $D$ represents the fraction of the input photons that contribute to the readout contrast \cite{danjou2019}, which we call the efficiency, effectively measuring the visibility of the dispersive shift against the resonator linewidth:
\begin{eqnarray}
    D =&& \frac{\kappa^2\chi_z^2}{[\left(\kappa/2\right)^{2}+\left(\delta_d - \chi_z\right)^2][\left(\kappa/2\right)^{2}+\left(\delta_d + \chi_z\right)^2]}.
\end{eqnarray}
$\langle n \rangle \approx 2\langle\Omega\rangle^2/\kappa$ is a good approximation when the drive is on or near resonance with the qubit-state shifted resonator frequency. As with Ref.~\cite{danjou2019}, we maximize contrast by optimizing $D$ via the drive frequency. As shown in Fig.~\ref{fig:fig2}, for $|\chi_z|/\kappa < 0.5$, $D$ peaks at resonance ($\delta_d=0$) but remains sub-unity. For $|\chi_z|/\kappa > 0.5$, $D$ saturates at unity at two optimal detunings $\delta_d = \pm \sqrt{\chi_z^2 - \left(\kappa/2\right)^2}$. Assuming optimal driving, $D$ depends only on the ratio $|\chi_z|/\kappa$. In the weak shift regime ($|\chi_z|/\kappa<0.5$), this reduces to:
\begin{eqnarray}
    \label{eq:Dfunction}D=\frac{16\left(\chi_z/\kappa\right)^2}{\left(1+4\left(\chi_z/\kappa\right)^2\right)^2},
\end{eqnarray}
which drops significantly for lower $\chi_z$.

\subsection{\label{sec:fom}Readout figures of merit}
We assess readout performance using the signal-to-noise ratio (SNR) and readout fidelity $F$. In this work, we make use of $\text{SNR}^2$, which can be directly inserted into our expression for readout fidelity. It is a function of readout contrast and accounts for the presence of measurement and amplification noise in the final signal through the measurement variance $\sigma_m^2$:
\begin{eqnarray}
    \text{SNR}^2 := \frac{|\Delta\beta_{\text{out}}(t)|^2}{\sigma_m^2} = \frac{\kappa\langle n \rangle D}{2\sigma_m^2}.
\end{eqnarray}
Assuming minimal added noise during readout, e.g. through the use of a quantum-limited amplifier such that $\sigma_m^2 = 2\gamma$ \cite{blais2021,clerk2010}, we can express $\text{SNR}^2$ in terms of the measurement rate $\Gamma_m$ and the net relaxation rate $\gamma$:
\begin{eqnarray}
    \label{eq:snr}
    \text{SNR}^2 = \frac{\Gamma_m}{\gamma},
\end{eqnarray}
where $\Gamma_m = \kappa\langle n \rangle D / 4$ \cite{danjou2019}. We note that only $\gamma$ is relevant here, rather than the total decoherence rate $\gamma_{\text{dec}}$ introduced in Sec.~\ref{sec:qubitrelax}. Because dispersive readout constitutes a measurement along the qubit's $\sigma_z$ axis, pure dephasing $\gamma_{\phi}$ does not drive transitions between qubit states during the readout process. $\text{SNR}^2$ is thus exclusively bounded by $\gamma$.

For a given SNR$^2$, the qubit state may be resolved within a measurement time $t_m$ when $t_m \sim 1/\Gamma_m$. However, if the qubit decays before measurement is complete, readout fails. Thus, for high-fidelity readout, the qubit relaxation rate sets a limit: $t_m \ll 1/\gamma \implies\gamma \ll \Gamma_m$. Furthermore, signal leakage must occur faster than decay ($\kappa > \gamma$). While larger $\kappa$ implies higher measurement rates, excessive broadening of the resonator reduces contrast between the shifted frequencies and increases Purcell decay; $\kappa$ must therefore balance fast readout with contrast preservation.

In the regime where the qubit state cannot be accurately determined in the time $1/\kappa$, $\gamma \ll \Gamma_m < \kappa$, single-shot readout fidelity is approximated by \cite{danjou2019,gambetta2007,gambetta2008,danjou2014}
\begin{eqnarray}
    \label{eq:fopt}
    F \approx 1-\frac{1}{2\text{SNR}^2}\ln\text{SNR}^2.
\end{eqnarray}
which implies that high-fidelity readout requires $\text{SNR}^2 \gg 1$. We adopt the following reference thresholds for Sec.~\ref{sec:readoutresults}: $F \ge 90\%$ ($\text{SNR}^2 \ge 13$), $F \ge 95\%$ ($\text{SNR}^2 \ge 36$), and $F \ge 99\%$ ($\text{SNR}^2 \ge 282$).

\subsection{\label{sec:para_res}Parameter restrictions}
Here, we define the imposed restrictions on system parameters to ensure the validity of our theoretical model and the feasibility of readout.

\subsubsection{\label{sec:regimes}Regime validity and system requirements}
First, projecting into the ground state manifold requires energy levels of the unperturbed Hamiltonian to be well-separated. As established in Sec.~\ref{sec:disH}, we assume 
\begin{eqnarray}\label{eq:escale}
\omega_0 > \omega_B, \omega_r,    
\end{eqnarray}
which ensure that the excited orbitals are well-separated from the Zeeman-split lower orbitals and large resonator Fock states. We also assume 
\begin{eqnarray}\label{eq:spinlike}
    |\omega_0 - \omega_B| \gg A/4
\end{eqnarray}
such that the qubit states are predominantly spin-like. 

Second, convergence of the Schrieffer-Wolff transformations require $2||V|| < \Delta_{\text{min}}$, where $||V||$ is a perturbative term operator norm and $\Delta_{\text{min}}$ is the smallest relevant energy gap~\cite{bravyi2011,blais2021}. Similarly, to safely neglect the transition terms $H_{\text{tr}}$ in Eq.~\eqref{eq:Htr}, we impose the stricter bound:
\begin{eqnarray}\label{eq:bound}
    \left(\frac{||H_{\text{tr}}||}{\Delta_{\text{tr}}}\right)^2 < 10^{-2}.
\end{eqnarray}
This ensures unwanted transition probabilities between eigenstates remain below $1\%$, validating the use of the dispersive Hamiltonian $H_\text{dis}$ to describe the effective dynamics~\cite{danjou2019}.

Finally, as mentioned in Sec.~\ref{sec:fom}, the photon-loss rate must exceed the qubit relaxation rate, $\kappa > \gamma$. While not a validity condition for the Hamiltonian itself, this is a necessary condition for readout visibility.

Parameters violating these constraints are excluded.
\subsubsection{\label{sec:nc}Critical photon numbers}
While $\text{SNR}^2$ scales with $\langle n\rangle$, the dispersive approximation fails beyond a critical photon number $n_c$. Each photon-dependent qubit-flip transition that we treat perturbatively has its own critical number, $n_{c,1}, n_{c,2}, \dots, n_{c,6}$, determined by system parameters and computable individually. When $\braket{n}$ approaches one of these, the probability of the associated transition can no longer be ignored, and related nonlinear effects become significant. To avoid this regime, we calculate each $n_{c,i}$ and choose a conservative number
\begin{eqnarray}
    n_c &&= \min_i\left(n_{c,i}\right)\\
    \label{eq:meanphoton}\langle n \rangle &&= 0.1 \times n_c,
\end{eqnarray}
motivated by previous studies showing that nonlinear effects can become physically relevant well below $n_c$~\cite{blais2004,blais2021}. Explicit forms for $n_{c,i}$ are derived in Appendix~\ref{sec:ncderivation}. This ensures validity by limiting the photon number relative to the specific transition most susceptible to breakdown at each parameter point. We reject all parameters where $\braket{n} < 1$.

\subsection{\label{sec:workregime}Working regime}
In this section, we define the working regime of our analysis. We focus on a realistic parameter set: $\omega_r = 2\pi \times 6.5\text{ GHz}$ and $\omega_B = 2\pi \times 6.6\text{ GHz}$ \cite{tosi2017,osika2022,boross2016}, and a spin-resonator detuning $\Delta_- = 100\ \text{MHz}$ ensures the system remains within the dispersive regime. We expect our results and conclusions to extend qualitatively to other realistic parameters. 

\begin{figure*}
{\includegraphics[width=1.0\textwidth]{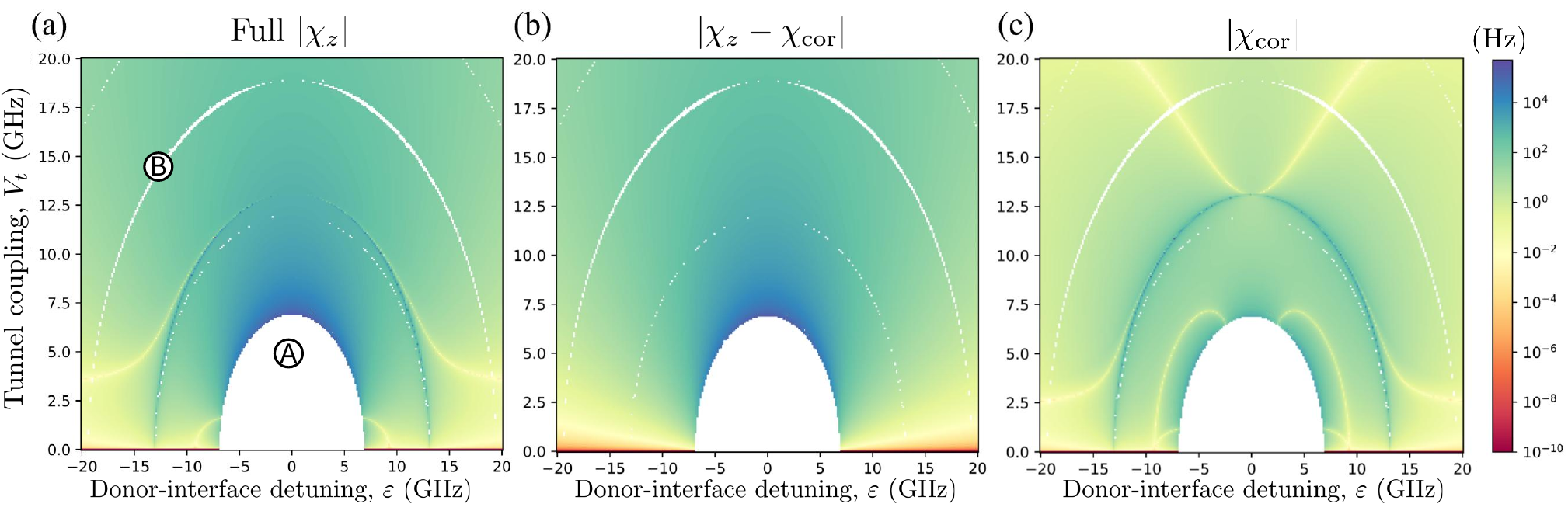}}
\caption{\label{fig:fig3} Using the parameters $g_c = 2\pi\times 50\text{ MHz}$, $\kappa = 2\pi\times 1.8\text{ MHz}$: (a) Full $|\chi_z|$ landscape over tunnel coupling $0\le V_t \le 2\pi\times 20\text{ GHz}$ and detuning $|\varepsilon| \le 2\pi\times 20 \text{ GHz}$. (b) Analytical approximation $|\chi_z-\chi_{\text{cor}}|$. (c) Magnitude of correction $|\chi_{\text{cor}}|$. Region A (white semi-ellipse) violate energy scales of Eq.~\eqref{eq:escale}, \eqref{eq:spinlike} or causes the dispersive approximation to break down ($\braket{n}<1$, Eq.~\eqref{eq:meanphoton}) from $n_{c,1}\rightarrow 0$.  Region B (white arcs) violate the unwanted transition bound of Eq.~\eqref{eq:bound}. Comparison of (c) with (a,b) confirms that $\chi_{\text{cor}}$ is negligible outside the regime where $\omega_0 \approx \omega_r + \omega_B$.}
\end{figure*}

Next, the expression for $\chi_z$ (Eq.~\eqref{eq:chi_1}) can be simplified to neglect $\chi_{\text{cor}}$. While our results later are obtained numerically, this makes their analysis and interpretation far simpler. Using the parameters $g_c = 2\pi\times 50\text{ MHz}$, $\kappa = 2\pi\times 1.8\text{ MHz}$ as an example, Fig.~\ref{fig:fig3} illustrates the validity of this approximation, which remain generally true.  Region A (white semi-ellipse) indicates the ``invalid region" where either the energy scales of Eq.~\eqref{eq:escale}, \eqref{eq:spinlike} fail or $n_{c,1}\rightarrow 0$ (Table.~\ref{tab:my-table2}), leading to the dispersive approximation breaking down ($\braket{n}<1$, Eq.~\eqref{eq:meanphoton}). Region B (white arcs) violates the unwanted transition bound of $H_{\text{tr}}$ in Eq.~\eqref{eq:bound}. We observe from Fig.~\ref{fig:fig3}(c) that the correction term $\chi_{\text{cor}}$ is negligible compared to $\chi_z-\chi_{\text{cor}}$, i.e. $|\chi_z - \chi_{\text{cor}}| \gg |\chi_{\text{cor}}|$, provided we operate away from frequencies $\omega_0 \approx \omega_r + \omega_B$. These correspond to spin-orbit transitions where the qubit's charge and spin degrees of freedom are simultaneously flipped~\cite{danjou2019}. This is analogous to the so-called straddling regime in superconducting qubits~\cite{koch2007,yamamoto2014,inomata2012,zotova2024,boissonneault2012}, which we exclude from our working space. We briefly discuss the regime's implications in Sec.~\ref{sec:straddling}.

We also adopt the approximations $\tilde{\omega}_r \approx \omega_r$, $\tilde{\omega}_{B} \approx \omega_B$, and $\tilde{\Delta}_- \ll \tilde{\Delta}_+$, since $\omega_B$ is in the neighborhood of $\omega_r$, which is a requirement for effective dispersive readout. Validated numerically, these simplify the dispersive shift to
\begin{eqnarray}
    \label{eq:chiapprox}
    \chi_z      &&\approx -\frac{g_{s}^{2}}{\Delta_-}.
\end{eqnarray}
This form is consistent with the standard RWA result, and similar approximations are used for the Purcell decay rate (Eq.~\ref{eq:purcell_rate}).

\section{\label{sec:readoutresults}Numerical results and discussion}

\subsection{\label{sec:swcoupling}Regions of high readout fidelity and strong coupling}
Our first objective is to characterize the qualitative trends of $\text{SNR}^2$, $\Gamma_m$, and $g_{s}$ across the working space defined by the detuning $\varepsilon$ and tunnel coupling $V_t$. Specifically, we investigate whether it is possible to simultaneously optimize these metrics to achieve both high single-shot readout fidelity and strong coupling. We define high fidelity as $F \ge 99\%$ ($\text{SNR}^2 \ge 282$) and strong coupling as $g_{s} \ge \text{max}\left(\kappa,\gamma_{\text{dec}}\right)$. Since the maximum decoherence rate in our working space is $\gamma_{\text{dec}} \approx 2\pi\times 0.26\ \text{MHz}$, the strong coupling condition effectively simplifies to $g_{s} \ge \kappa$ whenever $\kappa \gtrsim 2\pi\times 0.26\ \text{MHz}$.

\begin{figure*}
{\includegraphics[width=1.0\textwidth]{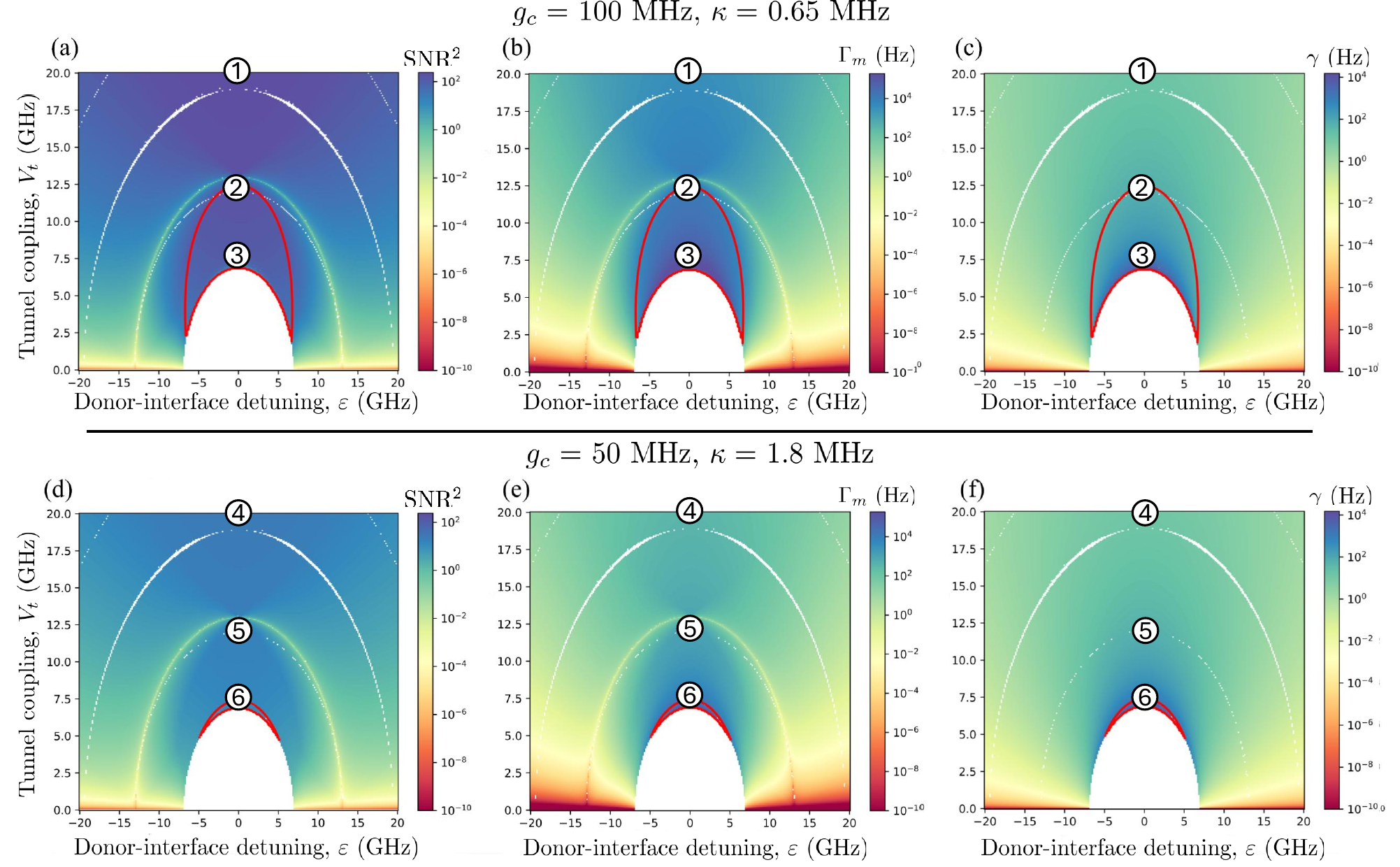}}
\captionof{figure}{\label{fig:fig4} Landscape of $\text{SNR}^2$, $\Gamma_m$ and $\gamma$ for $\omega_r = 2\pi \times 6.5 \text{ GHz}$ and $\omega_B = 2\pi \times 6.6 \text{ GHz}$. (a-c) Good parameters ($Q = 10^4$). (d-f) Sub-optimal parameters ($Q \approx 3.6\times 10^3$). We observe that high SNR$^2$ is concentrated at $\varepsilon = 0$ and high $V_t$. This is driven by the exponential suppression of the relaxation rate $\gamma$ from its maximum value of 0.012 MHz, which compensates for the reduction in measurement rate $\Gamma_m$. Conversely, $\Gamma_m$ peaks at lower $V_t$. The region circumscribed by the red contour indicates strong coupling ($g_{s}/\kappa \ge 1$), but this is only the small accessible part of the regime due to the invalid region, with the area being much smaller with sub-optimal parameters in (d-f). High $g_{s}/\kappa$ and $\Gamma_m$ can be simultaneously achieved, but the overlap between high SNR$^2$ and strong coupling is minimal. Maximizing SNR$^2$ typically requires operating in the weak coupling regime ($g_{s}/\kappa < 1$).}
\end{figure*}

Fig.~\ref{fig:fig4} maps the landscape for $\text{SNR}^2$, $\Gamma_m$, and $\gamma$ for two distinct parameter sets, representing good and sub-optimal experimental conditions:
\begin{enumerate}
    \item \textbf{Good (Fig.~\ref{fig:fig4}(a-c)):} $g_c = 2\pi\times 100\text{ MHz}$, $\kappa = 2\pi\times 0.65\text{ MHz}$, and a high quality factor $Q=10^4$ feasible with careful construction \cite{osika2022}.
    \item \textbf{Sub-optimal (Fig.~\ref{fig:fig4}(d-f)):} Reduced coupling $g_c = 2\pi\times 50\text{ MHz}$, higher loss $\kappa = 2\pi\times 1.8\text{ MHz}$, and lower $Q\approx3.61\times10^3$.
\end{enumerate}
Both chosen $g_c$ values are realistic for donor-based systems~\cite{tosi2017,osika2022}. The $\kappa$ values are realistic but were primarily chosen to highlight qualitative differences in results between the two sets. An exploration over the full $g_c$-$\kappa$ landscape is presented later in Sec.~\ref{sec:simultaneous}. The regions circumscribed by red contours indicate the strong coupling regime ($g_{s}/\kappa \ge 1$). For detailed numerical comparisons, Table~\ref{tab:my-table} lists the readout metrics for the specific points labeled 1-6 in Fig.~\ref{fig:fig4}.
\begin{table*}
    \centering
    \renewcommand{\arraystretch}{1.6}
    \begin{tabular}{|c||c|c|c|c|c|c|c|c|}
    \hline 
     \ & \textbf{Point} & $g_c/2\pi$~MHz & $\kappa/2\pi$~MHz & \textbf{$Q$} & $V_t/2\pi$~GHz & $\text{SNR}^2$ & $F$ &  $g_{s}/\kappa$\\ \hline\hline
\multirow{3}{*}{\rotatebox[origin=c]{90}{\text{Good}}} & 1 & \multirow{3}{*} {100} & \multirow{3}{*}{0.65} & \multirow{3}{*}{$10^4$} &  20  & 240 & 98.9\%  & 0.503 \\ \cline{2-2} \cline{6-9} 
\ & 2 &  &  &  &  12.3 & 182 & 98.6\% & 1.01 \\[5pt] \cline{2-2} \cline{6-9} 
\ & 3 &  &  &  &  7.3  & 55 & 96.5\% & 6.95\\ \hline
\multirow{3}{*}{\rotatebox[origin=c]{90}{\text{Sub-optimal}}} & 4 & \multirow{3}{*}{50} & \multirow{3}{*}{1.8} & \multirow{3}{*}{$3.61 \times 10^3$} & 20  & 25.8 & 93.7\% & 0.0909 \\ \cline{2-2} \cline{6-9} 
\ & 5 &  &  &  &  12.3 & 19.5 & 92.4\% & 0.182 \\ \cline{2-2} \cline{6-9} 
\ & 6 &  &  &  &  7.3  & 13.3 & 90.3\% &  1.25\\ \hline
    \end{tabular}
    \caption{Readout metrics and coupling strengths for points 1-6 in Fig.~\ref{fig:fig4}. We take $\varepsilon = 0$, $\omega_r = 2\pi \times 6.5 \text{ GHz}$, $\omega_B = 2\pi \times 6.6 \text{ GHz}$.}
    \label{tab:my-table}
\end{table*}
\subsubsection{General trends}
Fig.~\ref{fig:fig4}(a) and (d) reveal that $\text{SNR}^2$—and consequently the fidelity $F$—is highest at zero donor-interface detuning ($\varepsilon = 0$). This physically corresponds to where delocalization of the electron wavefunction and hence effective electric dipole are maximized. Transverse couplings leading to $g_s$, which drives readout through $\chi_z$, are enhanced, while longitudinal couplings that drive unwanted transitions are suppressed. This trend holds even as we extend into the invalid region. $\text{SNR}^2$ further improves as the tunnel coupling $V_t$ increases, peaking at 240 ($F = 98.9\%$) for the good parameters at point 1, which approaches but does not quite reach the high-fidelity threshold. 

Crucially, we observe a trade-off: while $\text{SNR}^2$ benefits from high $V_t$, both the measurement rate $\Gamma_m$ and the coupling ratio $g_{s}/\kappa$ are maximized at lower $V_t$, just outside the invalid region. This implies that simultaneous maximization of $\text{SNR}^2$ and coupling strength is not possible with these parameters. In fact, achieving the highest possible $\text{SNR}^2$, exceeding the range of $V_t$ shown, almost invariably forces the system into the weak coupling regime ($g_{s}/\kappa < 1$). The sub-optimal parameters in Fig.~\ref{fig:fig4}(d-f) show identical qualitative trends, albeit with globally reduced readout metric values and a significantly smaller strong-coupling window.
\subsubsection{Trends at zero detuning}
Motivated by the results above, we focus our analysis on the optimal regime at $\varepsilon = 0$ going forward, where our approximation for $\chi_z$ remains valid. To understand the behavior here, we must look at the interplay between the mean photon number $\langle n \rangle$, the readout efficiency $D$, and the relaxation rate $\gamma$ as $V_t$ varies, which are all non-linear and will generally speaking be non-monotonic. However, the condition $\omega_0>\omega_B,\omega_r$ simplifies things considerably. In many cases, it comes down to when the rate of change of $D$ dominates. 

As the orbital splitting $\omega_0$ increases with $V_t$, the mixing of the orbital with the spin and resonator subspaces is suppressed. This leads to a desirable reduction in both Purcell decay $\gamma_{\text{pu}}$ and phonon-mediated relaxation $\gamma_{\text{FF}}$ (see Fig.~\ref{fig:fig4}(c,f)), but it also causes the spin-photon coupling $g_{s}$ to drop. The readout efficiency $D$, which depends on $|\chi_z|/\kappa$, saturates at 1 when the coupling is sufficiently strong ($g_{s} \ge \sqrt{\kappa|\Delta_-|/2}$). However, in our accessible parameter space, this saturation condition only exists within the invalid region. Consequently, $D$ decreases monotonically as we increase $V_t$ according to Eq.~\eqref{eq:Dfunction}.

At $\varepsilon=0$, the mean photon number $\langle n \rangle$ is solely limited by the critical photon number associated with charge-resonator leakage into the excited state $\ket{e}$, i.e. the $\tau_x\left(a+a^{\dagger}\right)$ term:
\begin{eqnarray}
\braket{n}_{\varepsilon=0} = 0.1\times n_{c,1} = 0.1\times \frac{\Delta_0^2}{4g_c^2}
\end{eqnarray}
where $\Delta_0 = \omega_0-\omega_r$. Note that this expression is invalid and causes a discontinuity near resonances where $\omega_0\approx 2\omega_r$ or $\omega_0\approx \omega_r+\omega_B$ due to a breakdown of the dispersive approximation where $n_{c,4}$ or $n_{c,6}$ drop to 0 (Table~\ref{tab:my-table2}). It is also important that we use $n_{c,1}$ rather than the conventional Jaynes-Cummings limit $n_{c,5} = \Delta_-^2/4g_{s}^2$ \cite{blais2004,blais2021}. This is because $n_{c,1}$, which arises from the original Hamiltonian (Eq.~\ref{eqn:ffH}), is generally lower. Relying on $n_{c,5}$ would allow significant leakage into $\ket{e}$ due to photon pumping, thereby invalidating the dispersive approximation.

While $\braket{n}$ does increase with $V_t$ due to the larger energy gap, this boost is insufficient to counteract the rapid decay of the efficiency $D$ outside the immediate vicinity of saturation. As a result, $\Gamma_m$ is maximized at or near the low $V_t$ where $g_{s} = \sqrt{\kappa|\Delta_-|/2}$. Conversely, $\text{SNR}^2$ improves with $V_t$ because the exponential suppression of the relaxation rate $\gamma$ outweighs the drop in $\Gamma_m$. By extrapolating to $V_t > 2\pi\times 20\ \text{GHz}$, a maximum $\text{SNR}^2$ is expected to be found where the rate of $D$ and hence $\Gamma_m$ dropping starts to overtakes $\gamma$. This could theoretically allow $\text{SNR}^2$ to cross the $99\%$ threshold, but it would occur deeper in the weak coupling regime.

Qualitatively similar trends are observed for other sets of $g_c$ and $\kappa$. However, a distinction arises when $\kappa < 2\pi\times 0.26\ \text{MHz}$: in this limit, the strong coupling regime becomes defined by the decoherence rate $\gamma_{\text{dec}}$ rather than $\kappa$. Since $\gamma_{\text{dec}}$ follows an exponential decay with $V_t$, lower $\kappa$ values here effectively require a higher minimum $V_t$ to achieve strong coupling.

Additionally, for similar donor-based or quantum dot systems, the dominant spin relaxation process at high magnetic fields is often via phonon emission as well~\cite{tahan2014,amasha2008,hu2025}, analogous to $\gamma_{\text{FF}}$, so the same resulting trends are expected. However, the specific $\text{SNR}^2$ values they peak at and at what $V_t$ would depend on the system parameters present, at what point the rate of change in $D$ starts to take over, as well as on which transition whose critical photon number limits $\braket{n}$. 
\subsection{Regions of simultaneous high fidelity and strong coupling}
\label{sec:simultaneous}
Building on the analysis of $\text{SNR}^2$ and spin-photon coupling trends, we now seek to identify parameter regimes and system requirements where high readout fidelity ($F \ge 99\%$) and strong coupling ($g_{s} \ge \kappa$) coexist. While the specific parameters in Fig.~\ref{fig:fig4} did not reach the fidelity threshold, they indicate that simultaneously achieving both requires adjusting $g_c$ and $\kappa$ to extend the $V_t$ range for $g_{s}/\kappa\ge 1$ upwards and for $\text{SNR}^2\ge282$ downwards.

\begin{figure*}
{\includegraphics[width=1.0\textwidth]{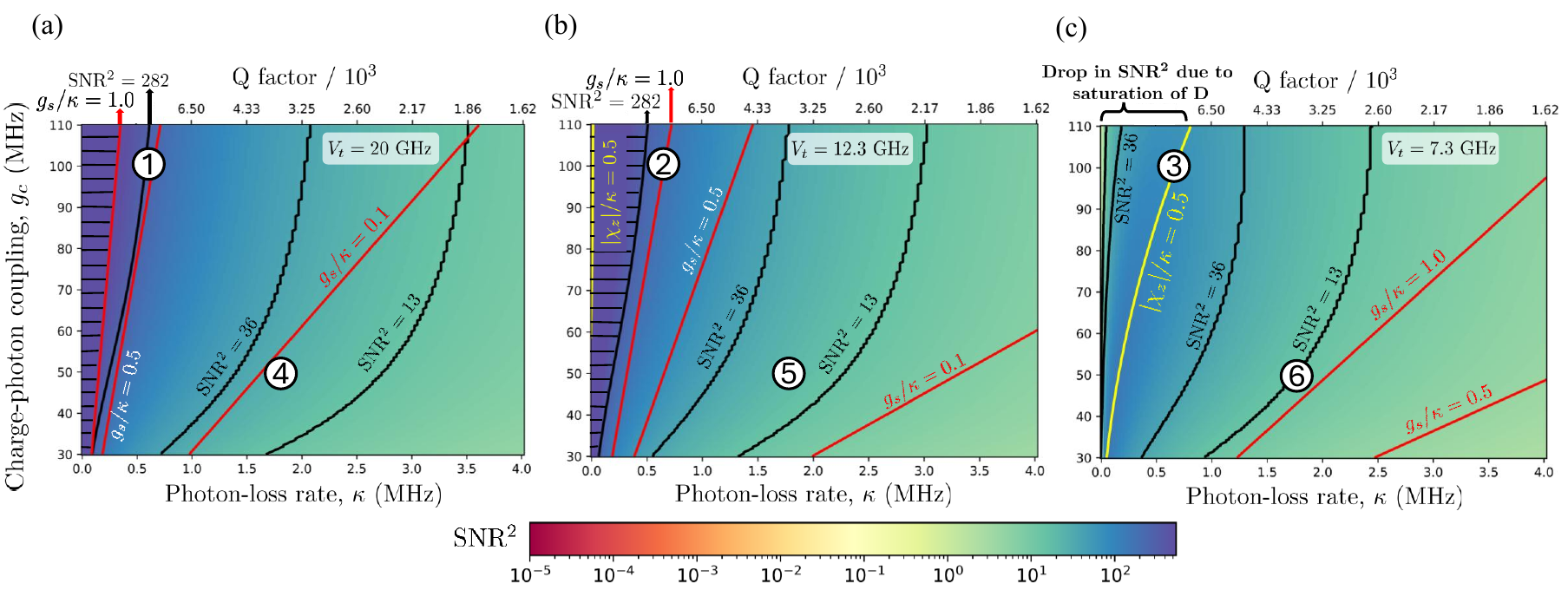}}
\caption{\label{fig:fig5} Regimes of simultaneous high fidelity and strong coupling. (a-c) Map of different $\text{SNR}^2$ (black contours: $F \ge 90\%$ ($\text{SNR}^2 \ge 13$), $F \ge 95\%$ ($\text{SNR}^2 \ge 36$), and $F \ge 99\%$ ($\text{SNR}^2 \ge 282$)) and $g_{s}/\kappa$ (red contours) thresholds for fixed tunnel couplings $V_t$. The minimum photon-loss rate of the plot is $\kappa = 2\pi \times6.5\ \text{kHz}$, and corresponding quality factors are given. Striped areas highlight the target regime where strong coupling ($g_{s}/\kappa \ge 1$) and high fidelity ($F \ge 99\%$) coexist. This area is maximized at intermediate tunnel couplings (comparing (b) to (a)). The yellow contour ($|\chi_z|/\kappa = 0.5$) demarcates the saturation of readout efficiency $D$. To the left of this line ($D=1$), increasing $g_c/\kappa$ yields no efficiency gain and instead degrades $\text{SNR}^2$ via increased Purcell decay and lowered mean photon number $\braket{n}$, as observed in (c). Numbered points 1-6 correspond to the specific parameter sets from Fig.~\ref{fig:fig4}.}
\end{figure*}

Fig.~\ref{fig:fig5} maps these thresholds across $g_c$ and $\kappa$ for the fixed $V_t$ values from Table~\ref{tab:my-table}. Since we operate at $\varepsilon=0$, the decoherence rate $\gamma_{\text{dec}} \approx \gamma_{\phi}$ is assumed constant for a given $V_t$. This sets a lower bound on $\kappa$; if $\kappa$ drops below $\gamma_{\text{dec}}$, the strong coupling condition would depend on $\gamma_{\text{dec}}$ rather than $\kappa$. However, for the contours shown, $\kappa$ remains the limiting factor. Contours with $\gamma_{\text{dec}}$ being the limiting factor would show an obvious kink. The condition $\kappa>\gamma$ also limits the allowed $g_c$ and $\kappa$ pairs due to $\gamma_{\text{pu}}$'s dependence on them, although this does not appear within our working space.

Our targets are the striped regions where strong coupling and high readout fidelity are met. The behavior of $\text{SNR}^2$ and $g_{s}/\kappa$ once again depend on the interplay of their different components. Unsurprisingly, increasing $g_c$ proportionately boosts $g_{s}$, while decreasing $\kappa$ lowers the threshold for strong coupling. This is evident when comparing Fig.~\ref{fig:fig5}(b) and (c): shifting from point 5 to point 2 (higher $g_c/\kappa$) enables strong coupling at a higher tunnel coupling ($V_t = 2\pi\times 12.3\ \text{GHz}$ vs $7.3\ \text{GHz}$). Increasing the $g_c/\kappa$ ratio thus effectively extends the accessible $V_t$ range for strong coupling.

To analyze $\text{SNR}^2$, we rewrite Eq.~\eqref{eq:snr} to isolate the dependence on $g_c^2/\kappa$:
\begin{eqnarray}
\label{eq:snr_rewritten}
    \text{SNR}^2=\frac{\Delta_0^2}{40}\frac{\kappa}{g_c^2}\times D\times \frac{1}{4\left(\gamma_{\text{FF}}+\kappa g_c^2 R\right)}
\end{eqnarray}
where $R = (g_{s}/g_c\Delta_-)^2$. Here, the first term represents $\kappa\braket{n}$, the second is the efficiency $D$ (which scales with $g_c^2/\kappa$), and the final term accounts for relaxation rates. $\gamma_{\text{FF}}$ is independent of $g_c^2/\kappa$, so the dependence there comes from $\gamma_{\text{pu}}$. 

When $D<1$, the rapid growth of efficiency $D$ with $g_c^2/\kappa$ dominates the SNR behavior. Consequently, optimal values lie near or on the yellow contour ($|\chi_z|/\kappa = 0.5$) where $D$ saturates. This is ideal, as $\text{SNR}^2$ growth at any specific $V_t$ implies an overall increase in $\text{SNR}^2$ across all $V_t$. Along this contour, since $D$ and $\kappa\braket{n}$ are constant, $\text{SNR}^2$ is maximized at lower $g_c$ and $\kappa$ where Purcell decay is minimized. Beyond this saturation point (Fig.~\ref{fig:fig5}(c)), increasing $g_c^2/\kappa$ yields no further efficiency gain but degrades SNR by increasing Purcell decay and reducing $\kappa\braket{n}$. Therefore, if a high-fidelity regime exists, the yellow contour will pass through it. Conversely, should high fidelity be absent from the entire contour (Fig.~\ref{fig:fig5}(c)), we can conclude that the regime is not possible for that $V_t$ for all $g_c$ and $\kappa$. Increasing $g_c/\kappa$ (and hence $g_c^2/\kappa$) is thus the key, up to the saturation limit, to extending the high-fidelity regime as well.

The resulting striped regions in Fig.~\ref{fig:fig5}(a,b) indicate viable parameters for the simultaneous regime. For example:
\begin{enumerate}
    \item At $V_t = 2\pi\times 20\ \text{GHz}$ ($\gamma_{\text{dec}} \approx 91.3\ \text{kHz}$), we require $\kappa \le 2\pi\times92.1\ \text{kHz}$ ($Q\ge7.1\times10^4$) for $g_c=30\ \text{MHz}$, or $\kappa \le 2\pi\times0.355\ \text{MHz}$ ($Q\ge1.83\times10^4$) for $g_c=110\ \text{MHz}$.
    \item At $V_t= 2\pi\times 12.3\ \text{GHz}$ ($\gamma_{\text{dec}} \approx 0.148\ \text{MHz}$), we require $\kappa\le 2\pi \times 80\ \text{kHz}$ ($Q\ge8.13\times10^4$) for $g_c=30\ \text{MHz}$, or $\kappa \le 2\pi\times0.517 \ \text{MHz}$ ($Q\ge1.3\times10^4$) for $g_c=110\ \text{MHz}$.
\end{enumerate}

Notably, the simultaneous regime is significantly larger in Fig.~\ref{fig:fig5}(b) than (a). This occurs because the $g_{s}/\kappa=1$ contour shifts left while the $\text{SNR}^2=282$ contour shifts right as $V_t$ increases. An intermediate $V_t$ where those contours align the most thus represents an optimal balance of $g_s$, $\gamma$, and $\braket{n}$, offering the greatest flexibility in $g_c$ and $\kappa$ to achieve simultaneous high fidelity and strong coupling. It also provides the widest window for $\text{SNR}^2$ and $g_{s}/\kappa$ values.

\begin{figure}
    {\includegraphics[width=0.50\textwidth]{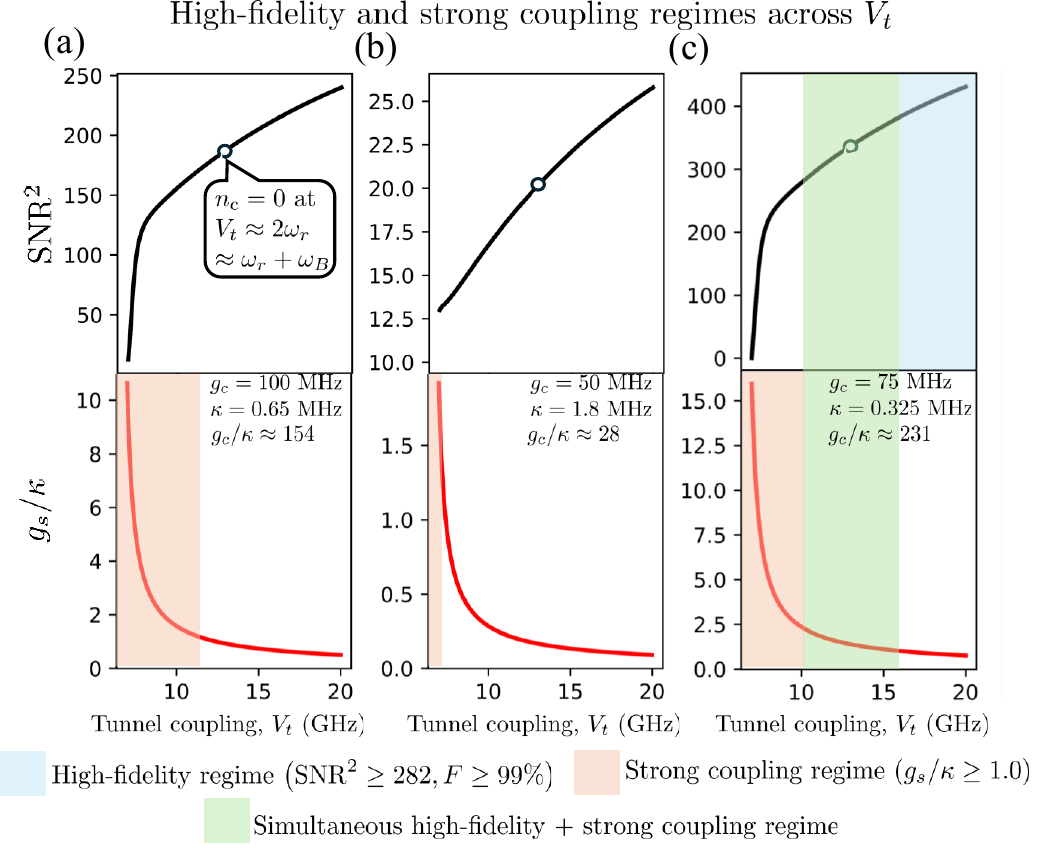}}
\caption{\label{fig:fig6} Comparison of high-fidelity and strong coupling regimes across $V_t$ for three parameter sets. (a) Good parameters from Fig.~\ref{fig:fig4}. (b) Sub-optimal parameters from Fig.~\ref{fig:fig4}. (c) New parameters ($g_c = 2\pi\times 75\text{ MHz}$, $\kappa = 2\pi\times 0.325\text{ MHz}$) selected from the overlapping region in Fig.~\ref{fig:fig5}(b). Shaded regions indicate: high fidelity ($\text{SNR}^2 \ge 282$, blue), strong coupling ($g_{s}/\kappa \ge 1$, red), and the simultaneous regime (green). The green overlap in (c) demonstrates that with adequate $g_c$ and $\kappa$, an intermediate tunnel coupling ($V_t \approx 10-16\ \text{GHz}$) allows for simultaneous high-fidelity readout and strong coupling. The discontinuity in the $\text{SNR}^2$ curves corresponds to $V_t \approx 2\omega_r\approx\omega_r+\omega_B$, where the dispersive approximation fails as $n_{c,4},n_{c,6} \to 0$.}
\end{figure}

To further illustrate the findings from Fig.~\ref{fig:fig5}, Fig.~\ref{fig:fig6} plots the high-fidelity and strong coupling regimes as a function of $V_t$. We compare the two parameter sets from Fig.~\ref{fig:fig4} against a third set ($g_c=2\pi\times 75\ \text{MHz}$, $\kappa=2\pi\times 0.325\ \text{MHz}$), selected based on the striped region identified in Fig.~\ref{fig:fig5}(b). Note that this chosen $\kappa$ remains above the maximum decoherence rate $\gamma_{\text{dec}}\approx 2\pi \times 0.26\ \text{MHz}$, ensuring that $\kappa$ remains the limiting factor for strong coupling.

As expected, Figs.~\ref{fig:fig6}(a) and (b) show $V_t$ ranges that satisfy strong coupling (shaded red) but fail to reach the high-fidelity regime (shaded blue). However, Fig.~\ref{fig:fig6}(c) reveals a clear overlap (shaded green), indicating a simultaneous regime within the range $V_t \approx 2\pi\times(10-16 \ \text{GHz})$. The only exception, which we ideally work away from, is the discontinuity at $V_t\approx 2\omega_r \approx \omega_B + \omega_r$, where the dispersive approximation breaks down due to critical photon numbers $n_{c,4}$ and $n_{c,6}$ approaching zero. These plots confirm that higher $g_c/\kappa$ ratios effectively extend the usable $V_t$ range. Combined with Fig.~\ref{fig:fig5}, this analysis delineates the precise parameter space and system requirements necessary to achieve simultaneous high fidelity and strong coupling.

This systematic approach to identifying simultaneous regimes through numerical results is applicable to similar quantum dot and donor systems. However, experimental feasibility remains bound by bottlenecks in attainable charge-photon coupling $g_c$ and resonator quality. Mapping the parameter space in this way clarifies the strict requirements for $g_c$ and $\kappa$ and demonstrates how selecting an intermediate $V_t$ maximizes the operating window. While we fixed $\omega_B$ and $\omega_r$ in this analysis, we expect similar qualitative trends across different frequencies. Finally, we note that techniques such as squeezing offer a potential route to circumvent these bottlenecks, which we discuss in Sec.~\ref{sec:squeezing}.
\subsection{\label{sec:squeezing}Enhance coupling strength through squeezing}
It has been demonstrated that squeezing one of the resonator mode quadratures provides an exponential enhancement to the spin-photon coupling strength ($g_{s} \to g_{s} e^r$) \cite{leroux2018,kam2024,qin2018,qin2024}, which can significantly improve readout performance. While this potentially expands the simultaneous high-fidelity/strong-coupling regimes, it concurrently enhances other couplings, thereby increasing the probability of unwanted transitions. In our numerical model, this manifests as a reduction in critical photon numbers ($n_{c,i} \to n_{c,i} e^{-2r}$). We therefore examine the changes in allowed mean photon number $\braket{n}$ under the restriction of Eq.~\eqref{eq:meanphoton}.

\begin{figure}
{\includegraphics[width=0.5\textwidth]{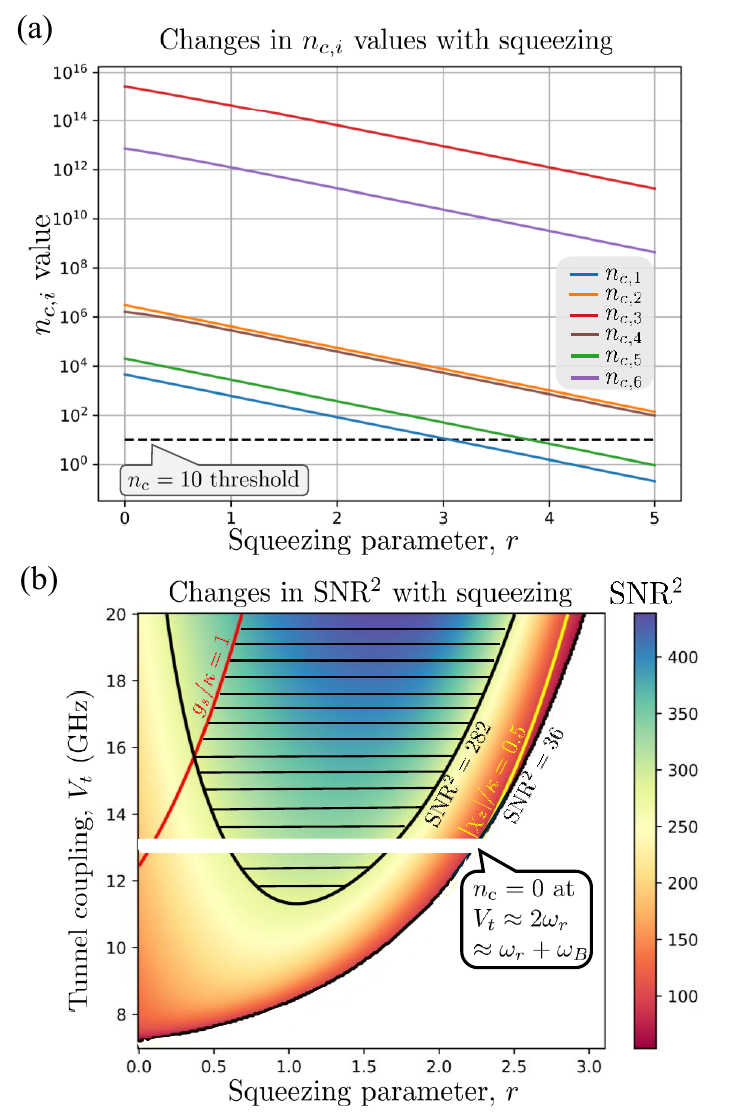}}
\caption{\label{fig:fig7} Effect of squeezing on critical photon numbers and readout performance. (a) Decay of critical photon numbers $n_{c,i}$ with squeezing parameter $r$ ($\varepsilon=0$, $V_t = 2\pi\times 20\ \text{GHz}$). Each line corresponds to a different $n_{c,i}$. The dashed line ($n_{c} = 10$) marks the chosen validity threshold: parameters become invalid for $r \gtrsim 3$ when $n_{c,1}$ crosses this line. (b) $\text{SNR}^2$ landscape against $r$ and $V_t$ (parameters from Fig.~\ref{fig:fig4}(a)). The white region indicates invalid parameters, with the discontinuity being the same one as Fig.~\ref{fig:fig6}. Black contours denote $\text{SNR}^2$ thresholds and the red contour marks the strong coupling boundary. The simultaneous regime (striped) is accessible with modest squeezing ($r \approx 0.37$) at intermediate tunnel coupling ($V_t \approx 2\pi\times15.6\ \text{GHz}$). With $r\gtrsim 0.7$, it is possible to work at $Vt = 2\pi\times 20\ \text{GHz}$. Unlike previous figures, the maximum $\text{SNR}^2$ ($\approx 439$, $F\approx 99.3\%$) does not align with the saturation of efficiency (yellow contour) but occurs at $r \approx 1.63$, after which $\text{SNR}^2$ declines due to reduced $\braket{n}$ and increased Purcell decay.}
\end{figure}

In Fig.~\ref{fig:fig7}(a), we plot the relevant $n_{c,i}$ values against the squeezing parameter $r$ for the good parameters of Fig.~\ref{fig:fig4}(a) at $\varepsilon = 0$ and $V_t = 2\pi\times20\ \text{GHz}$. Due to the large orbital splitting, we expect high $n_{c,i}$ but weak base $g_{s}$ here. We define an upper limit for $r$ where any $n_{c,i}$ drops below 10. In this case, $n_{c,1}$ limits the maximum squeezing to $r\approx 3$, corresponding to a 20-fold enhancement of $g_{s}$. However, this is still subject to experimental limitations for squeezing, which currently stand at $\sim 27\text{ db }\left(r \approx 3.10\right)$ for initial values and $\sim 15\text{ db }\left(r \approx 1.73\right)$ \cite{vahlbruch2016} for measured values after accounting for experimental losses. The latter corresponds to an enhancement of about 5.64 times.

Keeping the same parameters, Fig.~\ref{fig:fig7}(b) is a preliminary plot that maps out the readout landscape across a range of $V_t$ and $r$. Like earlier, we show the relevant contours, striped regions, and discontinuity. Note that we do not account for changes in measurement variance due to directional squeezing of the quadratures, and a full analysis to optimize those relevant phase mismatches \cite{kam2024} is beyond our scope. Consequently, squeezing here acts effectively as a boost to $g_c$, producing qualitative trends akin to Fig.~\ref{fig:fig5}. The results are promising: a simultaneous regime appears at intermediate tunnel couplings ($V_t \approx 2\pi\times 15.6\ \text{GHz}$) with only modest enhancement ($r\approx 0.37$). Furthermore, at $V_t=2\pi\times 20\ \text{GHz}$ where the maximum $\text{SNR}^2$ in this range lies, the simultaneous regime is accessible within $0.7\lesssim r\lesssim 2.3$, which falls comfortably within experimental limits. A wide range of $V_t$ values is also feasible for this intermediate range of $r$.

Crucially, the behavior of the maximum $\text{SNR}^2$ diverges from the trends in Sec.~\ref{sec:simultaneous}. Previously, with limited $g_c$ without squeezing, optimal SNR lay along the yellow contour where readout efficiency saturated ($D=1$). Squeezing, however, pushes the coupling strength far enough that efficiency gains from $D$ can no longer offset the increase in Purcell decay and reduction in $\braket{n}$. Further squeezing towards the saturation point would thus degrade $\text{SNR}^2$ despite still increasing $g_{s}/\kappa$, which creates a similar competing effect as described in Sec.~\ref{sec:swcoupling}. In Fig.~\ref{fig:fig7}(b), this limit manifests as a distinct peak in $\text{SNR}^2$ at $r\approx 1.63$ for $V_t=2\pi\times 20\ \text{GHz}$, achieving a maximum fidelity of $F\approx 99.3\%$—a threshold that was unattainable in the unsqueezed case ($r=0$). 

In summary, squeezing offers a viable pathway to overcome the $g_c$ bottleneck. With experimentally available levels of squeezing, the simultaneous regime is more easily reached across a wider range of system parameters. Modest squeezing is sufficient to yield mutual improvements in readout fidelity and coupling strength; however, one must identify the inflection point where $D$ loses its dominance in determining SNR behavior, as further squeezing yields diminishing returns for readout fidelity while continuing to enhance coupling. Further analysis is needed to understand if this remains the case when the full phase mismatches~\cite{kam2024} are involved. 
\subsection{\label{sec:additional}Special regions in parameter space}
While our analysis has focused on the $\varepsilon=0$ regime due to its optimal performance, we briefly discuss two notable regions at non-zero detuning ($\varepsilon \neq 0$) for completeness.
\subsubsection{\label{sec:sweetspot}Second-order clock transition sweet spot}
Beyond the first-order charge qubit sweet spot at $\varepsilon = 0$, the flip-flop qubit states themselves exhibit two first-order sweet spots. At specific coordinates ($\varepsilon,V_t$), they merge into a a second-order ``clock transition'' sweet spot (CTSS) \cite{tosi2017,truong2021}, where the qubit spin splitting becomes insensitive to charge noise to second order, presenting a potential working point. However, its proximity to (or overlap with) the invalid region renders it suboptimal. While the CTSS may still support strong coupling and fast readout, achieving high-fidelity likely requires enhancement techniques, such as squeezing. 
\subsubsection{\label{sec:straddling}Straddling regime}
\begin{figure}
{\includegraphics[width=0.5\textwidth]{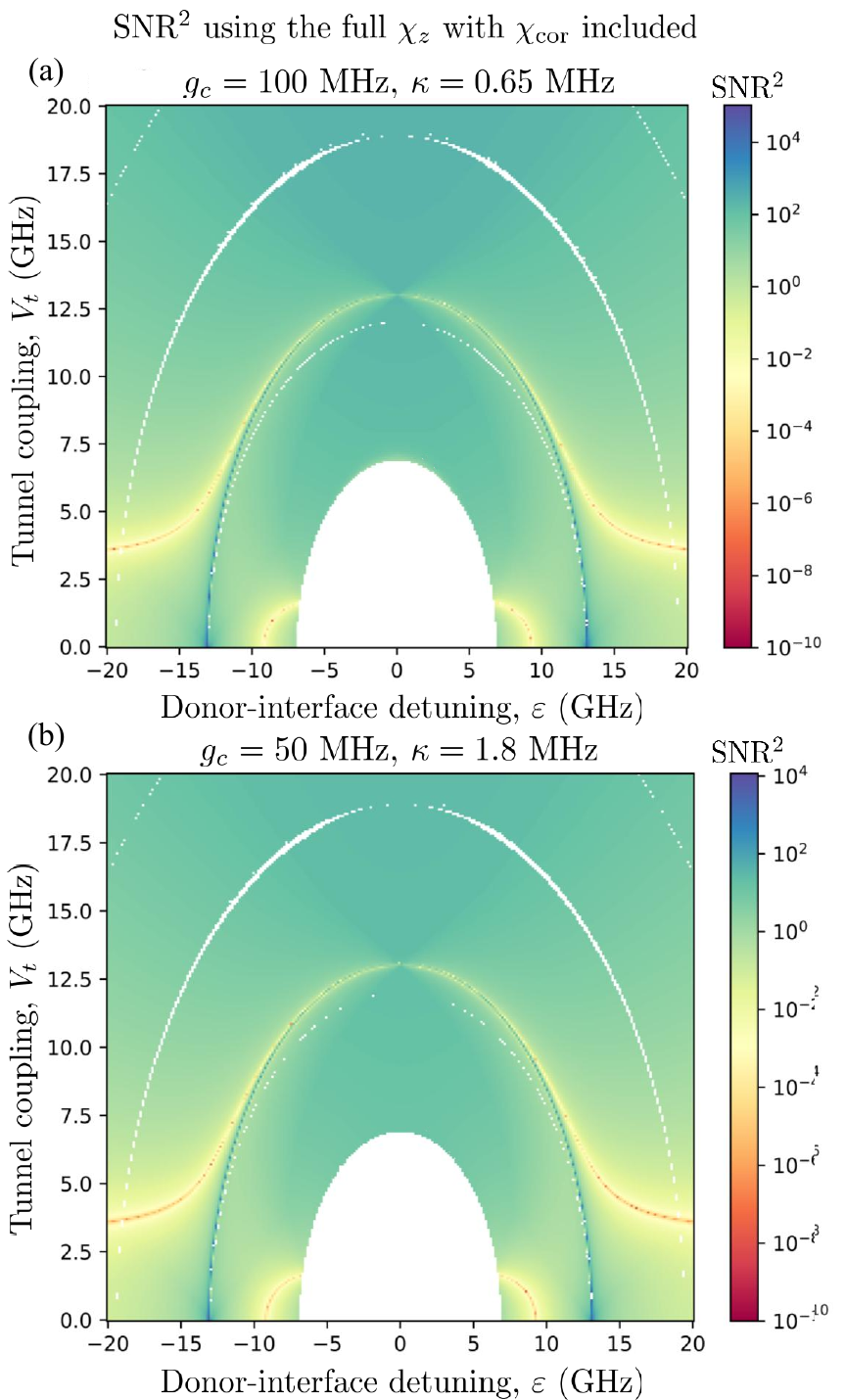}}
\caption{\label{fig:fig8} $\text{SNR}^2$ landscape calculated using the full $\chi_z$ (including $\chi_{\text{cor}}$) for the parameters of Fig.~\ref{fig:fig4}. Near the straddling regime ($\omega_0\approx\omega_r+\omega_B$), $\text{SNR}^2$ is significantly enhanced (dark blue) on one side and suppressed (red) on the other. While the enhanced region theoretically yields $\text{SNR}^2 > 1.0 \times 10^4$, the proximity to the resonance corresponds to a breakdown of the dispersive approximation where $n_{c,6} \to 0$ and parameters where $H_{\text{tr}}$ bounds are violated (white arcs near this regime), limiting the realistically feasible working range.} 
\end{figure}

In Sec.~\ref{sec:workregime}, we neglected $\chi_{\text{cor}}$ under the assumption of operating far from the straddling regime ($\omega_0\approx\omega_r+\omega_B$). Relaxing this assumption reveals that $\chi_{\text{cor}}$ diverges near this resonance. Depending on the sign of the detuning from this regime, $\chi_{\text{cor}}$ can constructively or destructively interfere with the dispersive shift, significantly enhancing or suppressing $\text{SNR}^2$~\cite{danjou2019,yamamoto2014,inomata2012,zotova2024,boissonneault2012}.
This effect is illustrated in Fig.~\ref{fig:fig8}. On the enhanced side of the straddling regime (dark blue), $\text{SNR}^2$ theoretically exceeds $1.0\times10^4$, surpassing the maximum values found in previous sections. However, realistic operation here is constrained because the resonance $\omega_0\approx\omega_r+\omega_B$ coincides with a breakdown of the dispersive approximation (vanishing $n_{c,6}$), as well as nearing parameters that violate $H_{\text{tr}}$ bounds (Eq.~\eqref{eq:bound}). Additionally, achieving strong coupling in this regime is considerably more difficult due to reduced mixing between the spin-orbital and orbital-resonator degrees of freedom. A comprehensive analysis of this regime is beyond the scope of this work.
\section{\label{sec:conc}Conclusions}
In this work, we have addressed the fundamental trade-off imposed by the spin-charge hybridization in scaling up semiconductor spin qubits via circuit QED. While hybridization is necessary to achieve the strong spin-photon coupling required for long-range interactions, it also introduces decoherence and relaxation channels that can compromise readout fidelity. 

Using the donor-based flip-flop qubit as a case study, we incorporated the valley-enhanced spin relaxation and critical photon numbers to rigorously map the interplay between these competing requirements. Our results demonstrate that a simultaneous regime of strong coupling and high fidelity is indeed achievable. We identified that the behavior of the SNR is primarily governed by the efficiency function $D$. The optimal operating points thus lie at intermediate tunnel couplings ($V_t \approx 10-16$ GHz in our example), which balance sufficient spin-charge mixing for strong coupling and preserving enough qubit lifetime to ensure high SNR and hence high readout fidelity. On the other hand, maximizing readout fidelity alone pushes the system towards the weak coupling regime, while maximizing coupling strength degrades fidelity due to enhanced relaxation. We further identified that the primary bottlenecks for accessing this simultaneous regime are the experimentally achievable charge-photon coupling ($g_c$) and photon-loss rate ($\kappa$) but found that squeezing can effectively mitigate these constraints. Modest squeezing can enhance the effective coupling strength enough to bridge weak and strong coupling regimes without requiring unrealistic device parameters, provided the deviation of trends at higher squeezing is accounted for.

Finally, while our analysis focused on the flip-flop qubit, the systematic approach applied here (mapping the boundaries of strong coupling against the efficiency-limited readout fidelity) is broadly applicable to any hybrid spin-charge system, including quantum dots and other donor-based architectures. 
\begin{acknowledgments}
This research is supported by the Ministry of Education, Singapore, under its Academic Research Fund Programme (T2EP50222-0017, RG154/24, and RG182/25) and National Research Foundation, Singapore and A*STAR under its Quantum Engineering Programme 2.0 (NRF2021-QEP2-02-P07). K. E. J. G. acknowledges funding from the National Research Foundation, Singapore through the National Quantum Office, hosted in A*STAR, under its Centre for Quantum Technologies Funding Initiative (S24Q2d0009). We acknowledge useful discussions with Susan Coppersmith. 
\end{acknowledgments}
\appendix
\section{\label{sec:SWderivation}Derivation of the effective dispersive Hamiltonian using the Schrieffer-Wolff transformation}
\subsection{First Schrieffer-Wolff transformation}
To derive an effective low-energy Hamiltonian that is needed for dispersive readout, we employ the Schrieffer-Wolff (SW) transformation, a form of degenerate perturbation theory, to decouple the low- and high-energy subspaces~\cite{schrieffer1966,bravyi2011}. The total Hamiltonian $H_{\text{tot}} = H_{\text{FF}} + H_{\text{r}} + H_{\text{int}}$ (see Sec.~\ref{sec:ff}) is decomposed as $H_{\text{tot}} = H_0 + V$, where the perturbation $V = V_{\text{od}} + V_{\text{d}}$ \cite{warren2019}, and
\begin{eqnarray}\label{eq:splittingH}
H_0 = &&\left(\frac{1}{8}A\cos{\eta}-\frac{1}{2}\omega_0\right)\tau_z+\frac{1}{8}A\sin{\eta}\tau_{x}\nonumber\\
&&+\left(-\frac{1}{2}\omega_B-\frac{1}{4}\Delta\omega_B\right)\sigma_z+\frac{1}{4}A\sigma_x\nonumber \\
&&-\frac{1}{8}A+\omega_ra^{\dagger}a\\
\label{eq:Vod}
 V_{\text{od}} =&&-\frac{1}{4}\Delta\omega_{B}\sin{\eta}\tau_{x}\sigma_{z}-\frac{1}{4}A\sin{\eta}\tau_{x}\sigma_{x}\nonumber \\
 &&+g_{c}\sin{\eta}\tau_{x}(a+a^{\dagger})  \\
  V_{\text{d}}=&&-\frac{1}{4}\Delta\omega_{B}\cos{\eta}\tau_{z}\sigma_{z}-\frac{1}{4}A\cos{\eta}\tau_{z}\sigma_{x}\nonumber\\
  &&+g_{c}\cos{\eta}\tau_{z}(a+a^{\dagger})
\end{eqnarray}
$H_{0}$ describes uncoupled dynamics within each orbital, spin, and resonator subspace; $V_{\text{od}}$ contains block off-diagonal terms inducing undesired orbital transitions, so we want to minimize these; and $V_{\text{d}}$ contains block-diagonal terms describing spin and resonator dynamics within each orbital. It is optimal to first diagonalize the orbital and spin subspaces of $H_0$ for the SW transformation, yielding
\begin{eqnarray}
H_0' = &&-\frac{1}{2}\omega_0'\tau_z'-\frac{1}{2}\omega_B'\sigma_z' + \omega_ra^{\dagger}a\\
 V_{\text{od}}' =&&\Delta\omega_{B,xz}\tau_{x}'\sigma_{z}'+A_{xx}\tau_{x}'\sigma_{x}' +g_{c,x}\tau_{x}'(a+a^{\dagger})  \\
V_{\text{d}}'=&&\Delta\omega_{B,zz}\tau_{z}'\sigma_{z}'+ A_{zx}\tau_{z}'\sigma_{x}'+g_{c,z}\tau_{z}'(a+a^{\dagger})
\end{eqnarray}
where now $V' = V_{\text{od}}'+V_{\text{d}}'$ and primed Pauli operators denote the new basis. The coefficients are:
\begin{eqnarray}
    \omega_0' =&& \frac{1}{4}\sqrt{A^2-8A\cos{\eta}\omega_0+16\omega_0^2}\\
    \tan\alpha =&& \frac{A\sin\eta}{-4\omega_0+A\cos\eta} \\
    \omega_B' =&& \frac{1}{2}\sqrt{A^2+4\omega_B^2+4\omega_B\Delta\omega_B+\Delta\omega_B^2} \\
    \tan\beta =&& \frac{A}{-2\omega_B-\Delta\omega_B} \\
    \Delta\omega_{B,xz} =&& -\frac{1}{4}\Delta\omega_B\left(\sin\eta\cos\alpha\cos\beta-\cos\eta\sin\alpha\cos\beta\right)\nonumber \\&&-\frac{1}{4}A\left(\sin\eta\cos\alpha\sin\beta-\cos\eta\sin\alpha\sin\beta\right) \\
    \Delta\omega_{B,zz} =&& -\frac{1}{4}\Delta\omega_B\left(\sin\eta\sin\alpha\cos\beta+\cos\eta\cos\alpha\cos\beta\right)\nonumber \\&&-\frac{1}{4}A\left(\sin\eta\sin\alpha\sin\beta+\cos\eta\cos\alpha\sin\beta\right) \\
    A_{xx} =&& -\frac{1}{4}\Delta\omega_B\left(-\sin\eta\cos\alpha\sin\beta+\cos\eta\sin\alpha\sin\beta\right)\nonumber\\-&&\frac{1}{4}A\left(\sin\eta\cos\alpha\cos\beta-\cos\eta\sin\alpha\cos\beta\right) \\
    A_{zx} =&& -\frac{1}{4}\Delta\omega_B\left(-\sin\eta\sin\alpha\sin\beta-\cos\eta\cos\alpha\sin\beta\right)\nonumber\\-&&\frac{1}{4}A\left(\sin\eta\sin\alpha\cos\beta+\cos\eta\cos\alpha\cos\beta\right) \\
    g_{c,x} =&& g_c\left(\sin\eta\cos\alpha-\cos\eta\sin\alpha\right) \\
    g_{c,z} =&& g_c\left(\sin\eta\sin\alpha+\cos\eta\cos\alpha\right) 
\end{eqnarray}
We perform the transformation $H' = e^{S}H_{\text{tot}}e^{-S}$ to second order in $V'$, with the generator $S$ chosen such that $[S,H_0' ]+V'_{\text{od}}  = 0$. Since $S$ is first order in $V'_{\text{od}}$, this eliminates $V'_{\text{od}}$ (block diagonalizes $H_{\text{tot}}$) to first order:
\begin{eqnarray}
    H' \approx H_{0}'+ V_{\text{d}}'+\frac{1}{2}\left[S,V_{\text{od}}'\right] + \left[S,V_{\text{d}}'\right]
\end{eqnarray}
The resulting Hamiltonian is:
\begin{eqnarray}
H'=&&\left(\omega_{r} +\omega_{r,x}\tau_x' +\omega_{r,z}\tau_z'\right)a^{\dagger} a \nonumber\\
&&+ \omega_{0(1)}\tau_z' + A_{0(1)}\tau_x' + \omega_{B(1)} \sigma^{'}_{z}  \nonumber\\
&&+A_{B(1)} \sigma^{'}_{x} + A_{zx}\tau_z'\sigma_x' + \Delta\omega_{B,zz}\tau_z'\sigma_z'\nonumber\\
&&+\left(g_{s,zx}\tau_z'\sigma_x'+g_{s,yy}\tau_y'\sigma_y'+g_{s,zz}\tau_z'\sigma_z'\right.\nonumber\\
&&\left.+g_{s,zI}\tau_z' + g_{s,xz}\tau_x'\sigma_z' + g_{s,xx}\tau_x'\sigma_x'\right)
\left(a + a^{\dagger}\right) \nonumber\\
&&+ \left(g_{s,yx}i\tau_y'\sigma_x'+g_{s,yz}i\tau_y'\sigma_z'\right)\left(a-a^{\dagger}\right)\nonumber \\
&&+ \left(g_{dou,z}\tau_z'+g_{dou,x}\tau_x'\right) \left(a^{2} + \left(a^{\dagger}\right)^{2}\right)\nonumber \\
&&+ g_{dou,y}i\tau_y'\left(a^2-\left(a^{\dagger}\right)^2\right)
\end{eqnarray}
where subscript $(1)$ denotes terms modified by the transformation. We retain all terms without projection at this juncture to preserve all interactions to second order. This only affects the final results at higher orders. 

\subsection{Second Schrieffer-Wolff transformation}
We rediagonalize the orbital and spin subspaces from $H'$, defining a new perturbation term ($V'' = V''_{\text{od}} + V''_{\text{d}}$) in the double-primed basis. The block off-diagonal component now causes spin-flip transitions:
\begin{eqnarray}
    H_0'' =&& -\frac{1}{2}\omega_{0(1)}'\tau_z''-\frac{1}{2}\omega_{B(1)}'\sigma_z''+\omega_ra^{\dagger}a\\
V_{\text{od}}'' =&&  \omega_{r,x}'\tau_x''a^{\dagger}a + A_{zx}'\tau_z''\sigma_x'' \nonumber\\
&&+\Delta\omega_{B,xz}'\tau_x''\sigma_z''+A_{xx}'\tau_x''\sigma_x''\nonumber\nonumber\\
&&+\left(g_{s,zx}' \tau_z''\sigma_x'' + g_{s,yy}\tau_y''\sigma_y'' \right.\nonumber\\
&&\left.+ g_{s,xz}'\tau_x''\sigma_z'' + g_{s,xx}'\tau_x''\sigma_x'' \right.\nonumber\\
&&\left.+ g_{s,xI}\tau_x''\right)
\left(a + a^{\dagger}\right) \nonumber\\
&&+ \left(g_{s,yz}'i\tau_y''\sigma_z'' + g_{s,yx}'i\tau_y''\sigma_x''\right)\left(a-a^{\dagger}\right)\nonumber\\
&&+ g_{dou,x}'\tau_x''\left(a^2+\left(a^{\dagger}\right)^2\right) \nonumber\\
&&+ g_{dou,y}i\tau_y''\left(a^2-\left(a^{\dagger}\right)^2\right)\\
V_{\text{d}}'' =&& \omega_{r,z}'\tau_z''a^{\dagger}a + \Delta\omega_{B,zz}'\tau_z''\sigma_z'' \nonumber \\
&&+ \left(g_{s,zz}'\tau_z''\sigma_z''+ g_{s,zI}'\tau_z''\right)\left(a+a^{\dagger}\right) \nonumber\\
&&+ g_{dou,z}'\tau_z'' \left(a^2+\left(a^{\dagger}\right)^2\right)
\end{eqnarray}
Of particular note are the coupling terms $g_{s,ij}$. The magnitude of the dominant spin-photon coupling term $|g_{s,zx}'|$ is our spin-photon coupling strength, used later to calculate the dispersive shift $\chi_z$ and the Purcell decay $\gamma_{\text{pu}}$. At little cost to accuracy, we can approximate it as
\begin{eqnarray}
    g_s = |g_{s,zx}'|\approx |g_{s,zx}| =&& \left|\frac{A_{xx} g_{c,x}}{2} \left(- \frac{1}{ \omega^{'}_{0} +  \omega_{r}} - \frac{1}{ \omega^{'}_{0} -  \omega_{r}}\right.\right. \nonumber\\
    &&\left.\left.- \frac{1}{ \omega^{'}_{0} +  \omega^{'}_{B}} - \frac{1}{ \omega^{'}_{0} -  \omega^{'}_{B}}\right)\right|
\end{eqnarray}
We now apply the second transformation $H_{\text{eff}} = e^{S'}H'e^{-S'}$ with generator $S'$ up to second order in $V''$ to decouple spin-flip transitions:
\begin{eqnarray}
    H_{\text{eff}} \approx H_{0}''+ V_{\text{d}}''+\frac{1}{2}\left[S',V_{\text{od}}''\right] + \left[S',V_{\text{d}}''\right]
\end{eqnarray}
This yields some third and fourth order terms in $V'$; since the first transformation was truncated at second order, these terms are inaccurate. However, we are only interested in an approximation up to second order, so dropping these higher order terms has a negligible effect on our results. Dropping additive constants as well, this gives us
\begin{eqnarray}
    \label{eq:Hdis}
    H_{\text{eff}} &&= H_{\text{dis}} + H_{\text{tr}}
\end{eqnarray}
where
\begin{eqnarray}
    H_{\text{dis}} =&& -\frac{1}{2}\omega_{B(2)}' \sigma_z'' + \left(\omega_{r(2)}'  + \chi_{z}\sigma_z''\right)a^{\dagger}a\\
    \label{eq:Htr}
    H_{\text{tr}} =&& \sum_{\left(i,j\right)} \left(\lambda_{ij}\tau_i \sigma_j +\xi_{ij}\tau_i \sigma_j a^{\dagger}a\right)\nonumber\\
    &&+ \sum_{\left(k,l\right)} \left(g_{sp,kl}\tau_k \sigma_l \left(a+a^{\dagger}\right) + g_{d,kl}\tau_k \sigma_l\left(a^2+\left(a^{\dagger}\right)^2\right)\right)\nonumber\\
    &&+ \sum_{\left(m,n\right)} i\left(g_{sp,mn}\tau_m \sigma_n \left(a-a^{\dagger}\right) \right.\nonumber\\
    &&\left.+ g_{d,mn}\tau_j \sigma_k\left(a^2-\left(a^{\dagger}\right)^2\right)\right)
\end{eqnarray}
Here, $H_{\text{dis}}$ is the dispersive Hamiltonian projected onto the orbital ground state, valid under the restrictions of Sec.~\ref{sec:regimes}. $H_{\text{tr}}$ contains higher-order terms describing unwanted transitions between eigenstates of $H_{\text{dis}}$, which are negligible to second order. Subscript (2) like before denotes terms modified by the second transformation. The indices run over $\left(i,j\right)\in\left\{\left(x,0\right),\left(0,x\right),\left(x,x\right),\left(x,z\right),\left(y,y\right)\right\}$, $\left(k,l\right)\in\left\{\left(0,0\right),\left(x,0\right),\left(0,x\right),\left(x,x\right),\left(0,z\right),\left(x,z\right),\left(y,y\right)\right\}$ and $\left(m,n\right)\in\left\{\left(0,y\right),\left(y,0\right),\left(x,y\right),\left(y,x\right),\left(y,z\right)\right\}$.

The dispersive shift $\chi_z$ is given by
\begin{eqnarray}
    \chi_z =&& g_{s}^{2}\left(-\frac{1}{\omega_{B(1)}'-\omega_r}-\frac{1}{\omega_{B(1)}'+\omega_r}\right)\nonumber\\
    &&+\chi_{\text{cor}}
\end{eqnarray}
where the correction term $\chi_{\text{cor}}$ (negligible in our parameter space, see Fig.~\ref{fig:fig3}) is
\begin{eqnarray}
    \label{eq:chicor}
    \chi_{\text{cor}}=&&\left(\frac{g_{s,xx}^2}{2}+\frac{g_{s,yx}^2}{2}+\frac{g_{s,yy}^2}{2}\right.\nonumber\\&&\left.+g_{s,xx}g_{s,yx}+g_{s,xx}g_{s,yy}+g_{s,yx}g_{s,yy}\right)\nonumber\\
    &&\cdot \left(- \frac{1}{\omega_{0(1)}' + \omega_{B(1)}' + \omega_{r}} \right.\nonumber\\&&\left.- \frac{1}{ \omega_{0(1)}' + \omega_{B(1)}' - \omega_{r}}\right. \nonumber\\
    &&\left.+ \frac{1}{ \omega_{0(1)}' -  \omega_{B(1)}' +  \omega_{r}} \right.\nonumber\\&&\left.+ \frac{1}{ \omega_{0(1)}' -  \omega_{B(1)}' -  \omega_{r}}\right).
\end{eqnarray}
In the main text, we use tilde notation for readability:
\begin{eqnarray}\label{eq:tildedefinitions}
    \tilde{\omega}_r\equiv \omega_{r(2)}', \quad
    \tilde{\omega}_B\equiv \omega_{B(2)}', \quad
    \tilde{\Delta}_\pm\equiv \omega_{B(1)}'\pm\omega_r.
\end{eqnarray}
Primes for the Pauli operators were also dropped at negligible accuracy loss.
\section{\label{sec:ncderivation}Derivation of the critical photon numbers}
The dispersive approximation requires the mean photon number $\braket{n}$ to be significantly lower than the critical photon numbers $n_{c,i}$ for all relevant qubit-photon transitions. This allows the transitions to be safely treated perturbatively. Here, we consider the transitions present in $V'_{\text{od}}$ and $V''_{\text{od}}$. Convergence of the SW transformation requires $2||V|| < \Delta_{\text{min}}$ \cite{bravyi2011,blais2021}. We can thus solve for $n$ from this inequality.

For an explicit example, consider the single-photon charge excitation channel $\ket{g,n} \leftrightarrow \ket{e,n-1}$ associated with $n_{c,1}$ and the term $g_{c,x}\tau_x\left(a+a^{\dagger}\right)$. The operator norm then involves finding the matrix element
\begin{eqnarray}
    ||V_{1}||&&=g_{c,x}\bra{e,n-1}\tau_x\left(a+a^{\dagger}\right)\ket{g,n}\nonumber\\
    &&= g_{c,x}\sqrt{n}
\end{eqnarray}
and the energy gap is $\omega_0'-\omega_r$. Thus,
\begin{eqnarray}
    n_{c,1} = \frac{\left(\omega_0'-\omega_r\right)^2}{4g_{c,x}^2}
\end{eqnarray}
Using this, Table~\ref{tab:my-table2} lists $n_c$ expressions for all relevant transitions. In our working space, we drop primes and number subscripts for clarity with negligible loss of accuracy.

\begin{table*}[h]

\centering
\renewcommand{\arraystretch}{3}
\begin{tabular}{|c|c|c|c|}
\hline
Label & \multicolumn{1}{c|}{$n_{c,i}$ Expression} & \multicolumn{1}{c|}{Associated Transition} & \multicolumn{1}{c|}{Effect on system when $\braket{n}\rightarrow n_{c,i}$} \\ \hline\hline
    $n_{c,1}$ & $\dfrac{\left(\omega_0'-\omega_r\right)^2}{4g_{c,x}^2}$ & $g_{c,x}\tau_x\left(a+a^{\dagger}\right)$ & \multicolumn{1}{c|}{\multirow{3}{*}{{\begin{tabular}[c]{@{}c@{}}Photon-induced leakage \\[-20pt] into orbital excited state $\ket{e}$ \end{tabular}}}}\\ \cline{1-3}
   $n_{c,2}$ & $\dfrac{\omega_{0(1)}'}{2\omega_{r,x}'}$ & $\omega_{r,x}'\tau_x a^{\dagger}a$ & \\ \cline{1-3}
  $n_{c,3}$ & $\dfrac{\left(\omega'_{0(1)}-\omega_r\right)^2}{4|g'_{s,xz}+g_{s,xI}+g'_{s,yz}|^2}$ & $\left(g'_{s,xz}\tau_x\sigma_z+g_{s,xI}\tau_x\right)\left(a+a^{\dagger}\right)+g'_{s,yz}i\tau_y\sigma_z\left(a-a^{\dagger}\right)$ & \\ \hline $n_{c,4}$ & $\dfrac{|2\omega_r-\omega'_{0(1)}|}{2|g_{dou,x}+g_{dou,y}|}$ & $g_{doux,}\tau_x\left(a^2+\left(a^{\dagger}\right)^2\right)+g_{dou,y}i\tau_y\left(a^2-\left(a^{\dagger}\right)^2\right)$ & \multicolumn{1}{c|}{\begin{tabular}[c]{@{}c@{}}Leakage into orbital excited state $\ket{e}$,\\[-20pt] squeeze-like dynamics from \\[-20pt] two-photon resonant exchange\end{tabular}}\\ \hline $n_{c,5}$ & $\dfrac{\left(\omega_{B(1)}'-\omega_r\right)^2}{4g_{s,zx}'^2}$ & $g_{s,zx}'\tau_z\sigma_x\left(a+a^{\dagger}\right)$ &  Unwanted photon-induced spin flip\\ \hline $n_{c,6}$ & $\dfrac{|\omega'_{0(1)}-\omega'_{B(1)}-\omega_r|^2}{4|g_{s,yy}+g'_{s,xx}+g'_{s,yx}|^2}$ & $\left(g_{s,yy}\tau_y\sigma_y+g'_{s,xx}\tau_x\sigma_x\right)\left(a+a^{\dagger}\right)+g'_{s,yx}i\tau_y\sigma_x\left(a-a^{\dagger}\right)$ &  \multicolumn{1}{c|}{\begin{tabular}[c]{@{}c@{}}Simultaneous spin-orbit flip \\[-20pt] (straddling regime) \end{tabular}}\\ \hline
\end{tabular}%
\caption{Expressions for critical photon numbers $n_{c,i}$, their associated transitions, and the physical effect on the system that causes the dispersive approximation to break down when $\braket{n}$ approaches each $n_{c,i}$. In our working space, we drop primes and number subscripts with negligible loss of accuracy.}
\label{tab:my-table2}
\end{table*}
%




\bibliography{sy_koh_references}

\end{document}